\documentstyle[11pt,fleqn,epsf]{article}
\renewcommand{\baselinestretch}{1.2}
\def\fnote#1#2{\begingroup\def\thefootnote{#1}\footnote{#2}\endgroup}
\makeatletter
\def\section{\@startsection {section}{1}{\z@}{3.5ex plus 1ex minus
    .2ex}{2.3ex plus .2ex}{\sc }}
\def\subsection{\@startsection{subsection}{2}{\z@}{3.25ex plus 1ex minus
   .2ex}{1.5ex plus .2ex}{\small \sc }}
\def\subsubsection{\@startsection{subsubsection}{2}{\z@}{3.25ex plus 1ex minus
   .2ex}{1.5ex plus .2ex}{\small \sc }}
\def\appendix{\par\clearpage
  \setcounter{section}{0}
  \setcounter{subsection}{0}
  \@addtoreset{equation}{section}
  \def\@sectname{Appendix~}
  \def\theequation{\thesection.\arabic{equation}}
  \def\thesection{\Alph{section}}}
\makeatother
\makeatletter \@addtoreset{equation}{section} \makeatother
\renewcommand{\theequation}{\thesection.\arabic{equation}}

\def\ap#1#2#3{     {\it Ann. Phys. (NY) }{\bf #1} (19#2) #3}

\def\npb#1#2#3{    {\it Nucl. Phys. }{\bf B #1} (19#2) #3}
\def\plb#1#2#3{    {\it Phys. Lett. }{\bf B #1} (19#2) #3}
\def\prd#1#2#3{    {\it Phys. Rev. }{\bf D #1} (19#2) #3}
\def\prep#1#2#3{   {\it Phys. Rep. }{\bf #1} (19#2) #3}
\def\prl#1#2#3{    {\it Phys. Rev. Lett. }{\bf #1} (19#2) #3}

\def\rmp#1#2#3{    {\it Rev. Mod. Phys. }{\bf #1} (19#2) #3}
\def\zpc#1#2#3{    {\it Z. Physik }{\bf C #1} (19#2) #3}

\def\nc#1#2#3{     {\it Nuovo Cim. }{\bf #1} (19#2) #3}

\def\ibid#1#2#3{   {\it ibid. }{\bf #1} (19#2) #3}
\def\ijmpa#1#2#3{  {\it Int. J. Mod. Phys. }{\bf A #1} (19#2) #3}
\def\eq#1{{eq.~(\ref{#1})}}
\def\eqs#1#2{{eqs.~(\ref{#1})--(\ref{#2})}}

\let\vev\VEV

\def\Tr{\mathop{\mbox{Tr}}\,}
\def\etal{{\it et al.}}

\newcommand{\bea}{\begin{eqnarray}}
\newcommand{\beq}{\begin{equation}}
\newcommand{\eea}{\end{eqnarray}}
\newcommand{\eeq}{\end{equation}}
\newcommand{\nnu}{\nonumber}
\newcommand{\spav}[1]{\parbox{1mm}{\vspace*{#1}}}

\begin{document}
\begin{titlepage}
\begin{flushright}
\vspace*{-1cm}
{\tt SISSA 43/95/EP}
\end{flushright}
\spav{0.0cm}
\begin{center}

{\large\bf The $\Delta S = 1$ Weak Chiral Lagrangian}\\
{\large\bf as the Effective Theory of the Chiral Quark Model}\\

\spav{0.8cm}\\
V. Antonelli$^{\dag\S}$, S. Bertolini$^{\S\dag}$, J.O. Eeg$^{\sharp}$,
M. Fabbrichesi$^{\S\dag}$ and
E.I. Lashin$^{\dag}$\fnote{\ddag}{Permanent address:
Ain Shams University, Faculty of Science, Dept. of Physics, Cairo, Egypt.}
\spav{1.2cm}\\
{\em  $^{\S}$ INFN, Sezione di Trieste.}\\

{\em $^{\dag}$ Scuola Internazionale Superiore di Studi Avanzati}\\
{\em via Beirut 4, I-34013 Trieste, Italy.}\\

{\em $^{\sharp}$ Department of Physics, University of Oslo}\\
{\em N-0316 Oslo, Norway.}\\
\spav{1.2cm}\\

{\sc Abstract}
\end{center}
We use the chiral quark model to construct the complete $O(p^2)$ 
weak $\Delta S = 1$ chiral lagrangian via the bosonization of
the ten relevant operators of the effective quark lagrangian.
The chiral coefficients are given
in terms of $f_{\pi}$, the quark and  gluon
condensates and the scale-dependent NLO Wilson coefficients of the
corresponding operators; in addition, they depend on  the
constituent quark mass $M$, a parameter characteristic of the model.
All contributions of order $N_c^2$ as well as
$N_c$ and $\alpha_s N_c$ are included.
The $\gamma_5$-scheme dependence of the chiral coefficients,
computed via dimensional regularization, and the Fierz transformation
properties of the operator basis are discussed in detail.
We apply our results to the evaluation of the hadronic matrix elements for the
decays $K \rightarrow 2 \pi$, consistently including the renormalization
induced by the meson loops. The effect of this renormalization
is sizable and  introduces a long-distance scale dependence that
matches in the physical amplitudes the short-distance scale
dependence of the Wilson coefficients.
\vfill
\spav{0.0cm}\\
{\tt SISSA 43/95/EP}\\
{\tt  September 1995 }

\end{titlepage}

\newpage
\setcounter{footnote}{0}
\setcounter{page}{1}

\section{Introduction}

Chiral perturbation theory
provides  a faithful representation of the
hadro\-nic sector of the standard model at low energies. The form of
this effective field theory is determined by $SU_L(3) \times SU_R(3)$
chiral invariance and its spontaneous breaking that,
together with Lorentz invariance,
 dictate all possible terms.

A particular example of
chiral perturbation theory is
the weak chiral lagrangian responsible for the $\Delta S = 1$ transitions.
This lagrangian controls most of the physics of kaon decays and in particular
the $\Delta I =1/2$ selection rule and the $CP$-violating
parameter $\varepsilon
'/\varepsilon$, the
determination of which is at the origin of this work.
To the  order $O(p^2)$, it is given
 in section~4. Such a lagrangian,
 by including also the effects of the
isospin-symmetry violating electromagnetic
interactions, is more general than that usually discussed
in the literature.

The coefficient in front of each term contains the short-distance
information. Since we do not know  how to connect directly
 the quark and gluon degrees of freedom of QCD to the hadronic states,
these coefficients have to  be determined either from a comparison to the
experiments or by means of some phenomenological model.
It is the purpose of this work to provide a
theoretical  determination of them
by means of the chiral quark model
($\chi$QM) approach~\cite{old,Cronin,QM} which is
briefly described in section 3.

In the $\chi$QM, and similar models, the interaction among mesons proceeds only
by means of quark loops (that explains why these models are often called
quark-loop models). Their theoretical justification can be found,
in terms of a more
fundamental picture, as the mean field approximation of an extended
Nambu-Jona-Lasinio (ENJL)
model~\cite{BBdeR} effectively mimicking QCD at intermediate energies.

Starting with the short-distance effective
$\Delta S = 1$ lagrangian (see section 2) written in terms of four-quark
operators, the $\chi$QM allows us to
compute the contribution
of each of these operators
to the corresponding coefficient of the low-energy chiral lagrangian. In the
process, all the arbitrary coefficients are parameterized in terms
of the input variables. These are  the Wilson
coefficients---that we take to the
next-to-leading order (NLO)---of the corresponding operators,
the pion decay constant $f_\pi$, the quark and
gluon condensates and  the constituent quark mass $M$,
a parameter characteristic of the model.

This determination requires some caution because of the presence of divergent
contributions. Regularization and renormalization prescriptions must
be given.
In order to be consistent with the regularization
of the anomalous dimensions of the short-distance analysis,
 we use
dimensional regularization in the modified minimal subtraction scheme.
In this framework, we must deal with the problems of
the $\gamma_5$-scheme dependence and of whether Fierz
transformations can be
applied to the original basis of chiral operators.
These problems are discussed in section 5, where we show
that the t'Hooft-Veltman (HV) scheme generates some fake chiral
anomalies that must be subtracted.

We show that
 operators related by a Fierz transformation lead to equivalent contributions
 to the chiral lagrangian coefficients in the HV scheme but not
 in the naive dimensional regularization (NDR) one. For the latter,
 Fierz  transformations with respect to the basis used for the
calculation of the Wilson coefficients must be avoided.
As an outcome, a
new $\gamma_5$-scheme dependence arises in the
coefficients of the chiral lagrangian in addition to  that, already
 present from the beginning,
of the NLO Wilson coefficients.
The matching between the two, which depends of the input
parameters, and the
resulting $\gamma_5$-scheme independence,  can be used to restrict the values
of
$M$, the only free parameter in the model. While such a procedure
is only mentioned in
passing here, it has been discussed in~\cite{BEF} and
 will be used extensively in future work.

In section 6, we write explicitly all the chiral lagrangian
coefficients determined in the HV
and NDR schemes. Both schemes are consistent for our lagrangian that contains
no anomaly. We have included contributions of order $O (N_c^2)$, $O (N_c)$
as well as $O (\alpha_s N_c )$. The latter represents non-perturbative
corrections induced by the gluon condensate; their relevance  was first
advocated
in~\cite{PdeR} where the effect on the operators $Q_1$ and $Q_2$ was
computed.

In  section 7, we discuss
the numerical values of the input parameters in terms of which these matrix
elements
are given by the $\chi$QM. The determination of these input values is
delicate and it requires some discussion because of the various estimates
presented in the literature.

For future applications, we specialize our results to
the decays $K^0 \rightarrow 2 \pi$ and give,
in section~8,
all the relevant hadronic matrix elements to order $O(\alpha_s N_c)$.  Such an
order includes the leading (that is $O(N_c^2)$) as well as the next-to-leading
order in the  $1/N_c$
expansion~\cite{1/N} and the leading corrections due to the gluon
condensate.  The reader who wishes to skip the technical
details
of how the result has been obtained is advised to go directly to this section.

 The  renormalization of the amplitudes induced by the meson loops
 (a correction of order $O(N_c)$) is sizable and it is given in  Section 8.1.
 Here our approach differs with respect to conventional chiral perturbation
theory in the way the scale independence of the physical amplitudes is
obtained.
 Whereas in the
conventional approach the scale dependence of the meson-loop
renormalization is compensated (by construction) by that of
the counterterms in the $O(p^4)$ weak chiral lagrangian,
in our approach these
counterterms are taken to be scale
independent at the tree-level. The chiral loops provide the
scale dependence of the hadronic matrix elements
that is  absent in the $\chi$QM itself.
It is through the matching
between this long-distance scale dependence and the short-distance one
contained in the Wilson coefficients  that the scale independence of the
physical
amplitudes should finally be achieved.

A similar point of view was first advocated and explored in
ref.~\cite{cutoff} (and subsequently in~\cite{Paschos})
in the framework of a cut-off regularization for the meson-loop renormalization
(while using dimensional regularization for the Wilson coefficients).
Our approach, based on dimensional regularization of the meson loops,
leads to different results, as discussed in section 8.

In the appendix,  the interested
reader can  find the  relevant Feynman rules and other formul\ae \  
useful in the
computation. We also briefly summarize the  HV and NDR $\gamma_5$-schemes and
give a table of the numerical value of all the input parameters.

\section{The $\Delta S = 1$  Effective Quark Lagrangian}

In our approach, the weak chiral lagrangian is determined by the
ten  operators   defining
 the $\Delta S = 1$ effective  lagrangian at scales
$\mu < m_c$~\cite{GW,Monaco}:
\beq
{\cal L}_{\Delta S = 1} =
- \frac{G_F}{\sqrt{2}} V_{ud} V_{us}^{*}  \sum_i \Bigl[
z_i(\mu) + \tau y_i(\mu) \Bigr] Q_i (\mu)
 \, . \label{ham}
\eeq
The $Q_i$ are four-quark operators obtained by integrating out in the standard
model the vector bosons and the heavy quarks $t,\,b$ and $c$. A convenient
basis is the following:
 \beq
\begin{array}{rcl}
Q_{1} & = & \left( \overline{s}_{\alpha} u_{\beta}  \right)_{\rm V-A}
            \left( \overline{u}_{\beta}  d_{\alpha} \right)_{\rm V-A}
\, , \\[1ex]
Q_{2} & = & \left( \overline{s} u \right)_{\rm V-A}
            \left( \overline{u} d \right)_{\rm V-A}
\, , \\[1ex]
Q_{3,5} & = & \left( \overline{s} d \right)_{\rm V-A}
   \sum_{q} \left( \overline{q} q \right)_{\rm V\mp A}
\, , \\[1ex]
Q_{4,6} & = & \left( \overline{s}_{\alpha} d_{\beta}  \right)_{\rm V-A}
   \sum_{q} ( \overline{q}_{\beta}  q_{\alpha} )_{\rm V\mp A}
\, , \\[1ex]
Q_{7,9} & = & \frac{3}{2} \left( \overline{s} d \right)_{\rm V-A}
         \sum_{q} \hat{e}_q \left( \overline{q} q \right)_{\rm V\pm A}
\, , \\[1ex]
Q_{8,10} & = & \frac{3}{2} \left( \overline{s}_{\alpha}
                                                 d_{\beta} \right)_{\rm V-A}
     \sum_{q} \hat{e}_q ( \overline{q}_{\beta}  q_{\alpha})_{\rm V\pm A}
\, ,
\end{array}  \label{Q1-10}
\eeq
where $\alpha$, $\beta$ denote color indices ($\alpha,\beta
=1,\ldots,N_c$) and $\hat{e}_q$  are the quark charges ($\hat{e}_d =
\hat{e}_s = - 1/3$ and $\hat{e}_u = 2/3$). Color
indices for the color singlet operators are omitted.
The subscripts $(V\pm A)$ refer to the chiral projections
$\gamma_{\mu} (1 \pm \gamma_5)$.
We recall that
$Q_{1,2}$ stand for the $W$-induced current--current
operators, $Q_{3-6}$ for the
QCD penguin operators and $Q_{7-10}$ for the electroweak penguins (and boxes).

The functions $z_i(\mu)$ and $y_i(\mu)$ are the
 Wilson coefficients evaluated at the scale $\mu$ and $V_{ij}$ the
Koba\-ya\-shi-Mas\-kawa (KM) matrix elements: $\tau = - V_{td}
V_{ts}^{*}/V_{ud} V_{us}^{*}$. 
Following the standard parameterization of the KM matrix, we have
that the $z_i(\mu)$ control the $CP$-conserving part of the amplitudes while
the
$y_i(\mu)$ the $CP$-violating one.

The numerical value of these Wilson coefficients depends on  $\alpha_s$, the
recent determination of which from NLO calculations on data taken at LEP and
SLC~\cite{alfa} gives
\beq
\alpha_s (m_Z) = 0.119 \pm 0.006 \, ,
\eeq
which corresponds to
\beq
\Lambda^{(4)}_{\rm QCD} = 350 \pm 100 \: \mbox{MeV} \, , \label{lambdone}
\eeq
the range of values we will use for numerical estimates.

Even though not all of these operators are independent, the basis in
\eq{Q1-10} has become
the standard one. As a matter of fact, it is the only one in which
the Wilson coefficients have been estimated to the
NLO order~\cite{Monaco,Roma}.

Two other possible operators, $Q_{11}$ and $Q_{12}$, representing the
dipole
contribution of the penguin operators, only contribute to the next order in
chiral perturbation and turn out to be negligible~\cite{BEF}.
On the other hand,
some of the other ten operators in \eq{Q1-10} may receive
sizable NLO corrections~\cite{EKW,Fej,BEF}.

\section{The Model}

In order to  evaluate  the bosonization
of the quark operators (\ref{Q1-10})
we exploit  the $\chi$QM approach
which provides an effective link between  QCD and
chiral perturbation theory.

The $\chi$QM  can be thought of as
the mean field approximation to the ENJL model of low-energy
QCD. A detailed discussion of the ENJL model and
its relationship with QCD---as well as with the $\chi$QM---
can be found, for instance, in ref.~\cite{BBdeR}. 

In the $\chi$QM, the light (constituent) 
quarks are coupled to the Goldstone mesons by  
the term
\beq
 {\cal{L}}_{\chi \mbox{\scriptsize QM}} = - M \left( \overline{q}_R \; \Sigma
q_L +
\overline{q}_L \; \Sigma^{\dagger} q_R \right) \, ,
\label{M-lag}
\eeq
where $q^T\equiv (u,d,s)$ is the quark flavor triplet, and
the $3\times 3$ matrix
\beq
\Sigma \equiv \exp \left( \frac{2i}{f} \,\Pi (x)  \right)
\label{sigma}
\eeq
contains the pseudoscalar octet
 $\Pi (x) = \sum_a \lambda^a \pi^a (x) /2 $,
$(a=1,...,8)$. The scale
$f$ is identified with the  pion decay constant $f_\pi$
(and equal to $f_K$ before chiral loops are introduced).

Under the action
of the generators $V_R$ and $V_L$ of the chiral
group $SU_R(3) \times SU_L(3)$, $\Sigma$ transforms linearly
\beq
\Sigma ' = V_R \Sigma V_L^{\dag} \, ;
\eeq
with the quark fields transforming as
\beq
q_L ' = V_L \, q_L \qquad \mbox{and} \qquad q_R ' = V_R \, q_R \, ,
\eeq
and accordingly for the conjugated fields.

The lagrangian (\ref{M-lag})
is invariant under the chiral group. If we consider the field
\beq
U = \widetilde{M} \Sigma
\eeq
as the exponential representation of the nonlinear sigma model, where
$\widetilde{M}$ is invariant under $SU(3)_V$ and $\Sigma$ is
given by eq.~(\ref{sigma}),
the $\chi$QM
can be understood as the mean field approximation in which $\widetilde{M} =
M \hat{I}$. The quantum excitations associated to
the full
$\widetilde{M}$ are thus decoupled in a manner that preserves chiral symmetry.
The quantity $M$, which is  characteristic of the model, can be interpreted
as the constituent quark mass that is generated, together with the quark
condensate, in the spontaneous  breaking of chiral symmetry.
It is given by
\beq
M = m_q - g \frac{\vev{\bar{q}q}}{\Lambda_\chi^2} \, ,
\eeq
where $g$ is a complicated function of the modeling of the solution
of the spontaneous breaking of chiral symmetry in QCD and
\beq
\Lambda_\chi \equiv 2 \pi \sqrt{\frac{6}{N_c}} f_\pi \,
\approx 0.82\ \mbox{GeV}.
\eeq
 In our approach we have
taken $M$ as a free parameter to be eventually constrained by
the $\gamma_5$-scheme independence of physical amplitudes (see the final
comment of section 8).

Fields other than the Goldstone bosons (like the $\rho$, for example) can in
principle be included by going beyond the mean field
approximation. They give  well-defined and computable
corrections~\cite{BBdeR}.

The gluon degrees of freedom of  QCD are considered as
integrated out down to the  chiral breaking  scale $\Lambda_\chi$, here  acting
as an infrared cut-off. The effect of the 
remaining soft gluons are assumed to be
well-represented by the various condensates, the leading contribution
coming from 
\beq
\langle \frac{\alpha_s}{\pi} G G \rangle \, .
\eeq
The constituent quarks are taken to be
propagating in the fixed background of such soft gluons.

 The $\chi$QM was first discussed
in~\cite{old,Cronin}. New life was breathed into it in a series of more recent
works~\cite{QM}.

We opted for the somewhat more restrictive definition
suggested in~\cite{EdeRT} (and there referred to as the QCD effective action
model) in which the meson degrees of freedom
do not propagate in the original lagrangian.

We have used the $\chi$QM of ref.~\cite{EdeRT} because:
\begin{itemize}
\item It is sufficiently simple without being trivial. It reproduces
important features we know from the experiments like, for instance,
the values of
the $O (p^4)$ strong chiral lagrangian coefficients $L_i$'s;

\item It is sensitive to the $\gamma_5$-scheme
dependence of dimensional regularization;

\item By freezing the meson degrees of freedom, it allows us to deal with
them in a separate step of the computation (see section 8).
\end{itemize}

The model interpolates between the chiral breaking scale
$\Lambda_{\chi}$ and $M$.
The constituent quarks are the only dynamical degrees of freedom
present within this range. The Goldstone bosons and
the soft QCD gluons are taken in our approach as external fields.
We neglect heavier scalar, vector and axial meson multiplets.

A kinetic term for the mesons, as well as the entire
chiral lagrangian
is generated and determined by
integrating out the constituent quark degrees of freedom of the model.
The $\Delta S =1$ weak
chiral lagrangian  thus becomes the effective theory of
the $\chi$QM
below the scale $M$. In the process, the many coefficients of the chiral
lagrangian are parameterized---to the order $O(\alpha_s N_c)$ in our
computation---in
terms of the input variables $f_\pi$,
the quark and gluon condensates and $M$, the only
free parameter of the model.

In the strong sector, where the
$O(p^4)$ coefficients $L_i$ are experimentally known, these can be used
to compare the prediction of the model. These coefficient have been
computed in~\cite{EdeRT}.
To the leading order in $N_c$,
the $L_i$ (except for $L_5$ and $L_8$) are purely
geometrical factors that cannot be compared directly with the experimental
values which have an explicit scale dependence. Only combinations of the same
that have vanishing anomalous dimension can be compared. In this case
the result is quite encouraging. See, for instance, $L_3$, $L_1 - L_2/2$ and
$L_9 + L_{10}$.
A more
stringent comparison can only be made by including the heavier multiplets that
are known to give rather large contributions to these coefficients. A recent
computation shows an impressive agreement~\cite{BBdeR}.

In the ENJL model  also
the other parameters of the strong sector---like, for instance,
$M$ and $f_\pi$---are in
principle
computable quantities themselves.
As attractive as this line of investigation might be, we
do not pursue it here. In our approach, the strong sector
parameters are only taken as input values, determined from
the experiments, in
order to predict the physics of the weak sector.

We would like to emphasize here the novelty of our approach with respect to the
usual treatment of the weak chiral lagrangian. This is best understood by
comparison
with the strong sector where the $\chi$QM uses the value of $f_\pi$ as an
input.
In this case the coefficient of the chiral lagrangian is scale independent and
there is no matching to any independently determined Wilson coefficient.
Because
of this, the strong sector follows the usual treatment and in particular
the higher-order counterterms must have a scale dependence in order to cancel
that
induced by the chiral loops. This is done by construction.
The weak sector behaves differently. The $\chi$QM
determines
the coefficients as a function of the Wilson coefficients so that they have
an explicit scale dependence that should then be matched to that independently
induced by the chiral loops. Whether this happens
represents a crucial test of our approach.
In this case the $O(p^4)$ chiral counterterms need not
have any scale dependence at the tree level.

As it is the case for the counterterms $L_i$ in the strong sector,
the effect of
the heavier multiplets on the $\chi$QM approximation may be important. In
particular, it is known~\cite{BBdeR} that the vector multiplets mix with the
Goldstone bosons thus modifying the axial coupling $g_A$ between constituent
quarks
and mesons that in the $\chi$QM is taken to be 1.
Their inclusion is postponed to future work.

\subsection{Bosonization}

A simple procedure allows us to find the
bosonization of the four-quark operators in \eq{Q1-10}.
 First, we
reproduce their $SU(3)$ flavor structure by using the appropriate
combinations of Gell-Mann matrices
acting on the quark flavor triplets $q$.
For instance, left-handed current operators (like $Q_1$ and $Q_2$)
are written as
\beq
\bar{q}_L \lambda^m_n\gamma^\mu q_L\ \ \bar{q}_L \lambda^p_q\gamma_\mu q_L \ .
\label{step1}
\eeq
where $\lambda^m_n$ ($m,n=1,2,3$) are the appropriate flavor projectors.
Then
we rotate the quark multiplets as follows:
\bea
q_L & = & \xi^{\dag} Q_L\, , \qquad \bar{q}_L = \bar{Q}_L \xi \nnu \\
q_R & = & \xi Q_R \, , \qquad \bar{q}_R = \bar{Q}_R \xi^{\dag} \, ,
\label{rotation}
\eea
where $\xi^2 = \Sigma$ and $(\xi^{\dag})^2 = \Sigma ^{\dag}$. The $Q_{L,R}$
are the constituent quark fields.

The rotation (\ref{rotation}) is such as
to transform the lagrangian of \eq{M-lag}
into a pure mass term for the constituent quarks, 
\beq
 - M \left(\overline{Q}_R Q_L + \overline{Q}_L Q_R \right) \, ,
\label{CQM}
\eeq
while the quark coupling to the Goldstone bosons is
transferred to the
kinetic term in the QCD lagrangian. The same rotation,  applied to the
operators in \eq{step1}, together with
their Fierzed transformed expressions,  yields
\beq
\bar{Q}_L \xi\lambda^m_n\gamma^\mu \xi^{\dag} Q_L\ 
\bar{Q}_L \xi\lambda^p_q\gamma_\mu \xi^{\dag} Q_L\quad 
\mbox{and}\quad (n\leftrightarrow q)\ .
\label{step2}
\eeq

Finally, the rotated quark fields are integrated out after having attached
the axial fields
\beq
A_\mu = - \frac{i}{2} \, \xi \Bigl( D_\mu \Sigma ^{\dag} \Bigr) \xi =
\frac{i}{2} \, \xi^{\dag} \Bigl( D_\mu \Sigma \Bigr) \xi^{\dag}
\eeq
to the quark loops in all possible manners that are consistent with
Lorentz invariance.
In the case of \eq{step2}, the insertion of no axial fields (constant term)
gives no contribution since
\beq
\Tr (\xi\lambda^m_n\xi^{\dag}\ \xi\lambda^p_q\xi^{\dag}) =
\Tr (\lambda^m_n\lambda^p_q) = 0\ ,
\label{constant}
\eeq
unless $(p,q)=(n,m)$. The first non-vanishing contributions are proportional
to
\bea
\Tr (\xi\lambda^m_n\xi^{\dag}\ A^\mu\ \xi\lambda^p_q\xi^{\dag}\ A_\mu) 
& \propto &
\Tr \Bigl(\lambda^m_n \ \Sigma^{\dag} D_\mu \Sigma\
          \lambda^p_q \ \Sigma^{\dag} D_\mu \Sigma \Bigr) 
          \nnu \\ 
\Tr (\xi\lambda^m_n\xi^{\dag}\ \xi\lambda^p_q\xi^{\dag}\ A_\mu A^\mu ) 
& \propto &
\Tr \Bigl(\lambda^m_n 
          \lambda^p_q  \ \Sigma^{\dag} D_\mu \Sigma\ 
          \Sigma^{\dag} D_\mu \Sigma \Bigr)  
          \label{pquadro} \\ 
\Tr (\xi\lambda^m_n\xi^{\dag}\ A^\mu A_\mu \ \xi \lambda^p_q\xi^{\dag}) 
& \propto &
\Tr \Bigl( \lambda^p_q \lambda^m_n \ 
           \Sigma^{\dag} D_\mu \Sigma \
          \Sigma^{\dag} D_\mu \Sigma \Bigr) \nnu
\eea
which provide, together with the $(n\leftrightarrow q)$ expressions,
the $O(p^2)$ chiral representation of the quark operator.
The single trace of \eq{pquadro} can be written 
as the
product of two traces (see  appendix A.4); in this form, the independent
terms are more easily recognized and it is easier to verify that the bosonized
forms satisfy the $CPS$ invariance~\cite{CPS} of the original quark operator.

The bosonization of operators involving the right-handed currents 
proceeds, along similar lines. Starting from the rotated operator
\beq
\bar{Q}_L \xi\lambda^m_n\gamma^\mu \xi^{\dag} Q_L\ \
\bar{Q}_R \xi^{\dag}\lambda^p_q\gamma_\mu \xi Q_R \ ,
\label{LRops}
\eeq
or  its Fierzed expression
\beq
\bar{Q}_L \xi\lambda^m_q\xi Q_R\ \
\bar{Q}_R \xi^{\dag}\lambda^p_n\xi^{\dag} Q_L\ ,
\label{LRopsF}
\eeq
we obtain after the identification of equivalent terms
\beq
\begin{array}{l}
  \Tr  \left( \lambda^m_n \ \Sigma^{\dag} \ \lambda^p_q \ \Sigma \right)\ , \\
 \Tr \left( \lambda^m_n \ D_\mu \Sigma^{\dag}\ 
\lambda^p_q \ D^\mu \Sigma \right)\ , \\
 \Tr  \left( \lambda^m_q \ \Sigma D_\mu \Sigma^{\dag} \ 
\lambda^p_n \ \Sigma^{\dag} D^\mu \Sigma \right)\ , \\
 \Tr  \left( \lambda^m_n  \ \Sigma^{\dag} \ \lambda^p_q \ D_\mu \Sigma 
 D^\mu \Sigma^{\dag}\ \Sigma \right) \ .
 \end{array}
\label{strano}
 \eeq
 
Not all of these bosonizations are actually present for each operator. 
For instance, in the case of 
the gluonic
penguins, the sum over the quark flavors together with unitarity make
all but one of these terms vanish. It is only for some of 
 the electroweak penguins that all contributions 
are actually there. 

The contributions arising from the 
last term in \eq{strano} are usually not
included in the literature (more on that in section 8.2). 
As a matter of fact,
by considering the $O(p^2)$ bosonization of
 the quark  density 
\beq
\bar{q}_L q_R  \to \frac{\vev{\bar q q}}{2}
\left[ \ \Sigma - \frac{c_1}{\Lambda_\chi^2} \ D_\mu D^\mu \Sigma -
\frac{c_2}{\Lambda_\chi^2}\  \Sigma\ D_\mu D^\mu \Sigma^\dag\ \Sigma
\right]\ ,
\label{qLqR}
\eeq
the neglect of the last term in \eq{strano} corresponds 
to discarding the second one of the two quadratic 
terms~\fnote{\dag}{A third term 
$c_3 \: D_\mu \Sigma D^\mu \Sigma ^{\dag}\ \Sigma$ 
can also be included. The three terms are however linearly dependent and
it is sufficient to keep only two of them as in \eq{qLqR}.
The equivalence of any choice of two terms can be explicitly verified. 
To this end it is
important to remember that the term proportional to $c_3$ induces a
wave function renormalization that must be included in
the determination of the coefficients, see section 8.2.}. 
In the first paper of ref.~\cite{1/N} the authors claim that
the constant $c_2$ ($c_1$) can be put to zero by a nonlinear 
transformation that
preserves the unitarity of $\Sigma$. We have verified that
such a transformation has no consequences in the case of the gluon Penguins,
where all $O(p^2)$ terms lead to the same bosonization
(proportional to the combination $c_1 + c_2$), while it cannot be applied
in the electroweak sector, where non-equivalent bosonizations are generated by 
the two independent terms (see section 8.2).

\section{The Weak Chiral Lagrangian}

The strong chiral lagrangian is completely fixed to the leading order in
momenta by symmetry requirements and the Goldstone boson's decay
amplitudes:
\beq
{\cal L}_{\rm strong}^{(2)} =
\frac{f^2}{4} \Tr \left( D_\mu \Sigma D^\mu \Sigma^{\dag} \right)
+ \frac{f^2}{2}
B_0 \Tr \left( {\cal M} \Sigma^{\dag} +  \Sigma {\cal M}^{\dag} \right) \, ,
\label{L2strong}
\eeq
where ${\cal M} = \mbox{diag} [ m_u, m_d, m_s ]$ and $B_0$ is defined by
$\langle \bar{q}_i q_j \rangle = - f^2 B_0 \delta_{ij}$.
To the next-to-leading order there are ten coefficients
$L_i$ to be determined~\cite{GL} .

For the weak $\Delta S =1$ flavor changing interactions,
the systematic application of the procedure described
in the previous section to the operators in \eq{Q1-10}
leads, after some algebraic manipulations, to the following
$O ( p^2 )$ chiral lagrangian:
\bea
{\cal L}^{(2)}_{\Delta S = 1}  = & &
G^{(0)}(Q_{7,8})
\Tr \left( \lambda^3_2 \Sigma^{\dag} \lambda^1_1 \Sigma
\right) \nnu \\
& +& G_{\underline{8}} (Q_{3-10}) \Tr \left( \lambda^3_2 D_\mu \Sigma^{\dag}
D^\mu \Sigma
\right)   \nnu \\
& +& \: G_{LL}^a (Q_{1,2,9,10}) \, 
\Tr \left(  \lambda^3_1 \Sigma^{\dag} D_\mu \Sigma \right)
\Tr \left( \lambda^1_2 \Sigma^{\dag} D^\mu  \Sigma \right)  \nnu \\
& +& \: G_{LL}^b (Q_{1,2,9,10})\, \Tr \left( \lambda^3_2 \Sigma^{\dag} D_\mu
\Sigma \right)
\Tr \left(  \lambda^1_1 \Sigma^{\dag} D^\mu \Sigma \right) \nnu \\
& +& \: G_{LR}^a (Q_{7,8})\, \Tr \left( \lambda^3_1 
D_\mu  \Sigma \right)
\Tr \left(  \lambda^1_2  D^\mu \Sigma^{\dag} \right)  \nnu \\
& +& \: G_{LR}^b (Q_{7,8})\, \Tr \left( \lambda^3_2 \Sigma^{\dag} D_\mu
\Sigma \right)
\Tr \left(  \lambda^1_1 \Sigma D^\mu \Sigma^{\dag} \right)  \nnu \\
& +& \: G_{LR}^c (Q_{7,8}) \left[ \Tr \left( \lambda^3_1 \Sigma \right)
\Tr \left( \lambda^1_2  D_\mu
\Sigma^{\dag} D^\mu \Sigma\ \Sigma^{\dag} \right) \right. \nnu \\
&  &  \quad \quad \quad \quad \quad + \left.
\Tr \left( \lambda^3_1  D_\mu \Sigma D^\mu \Sigma^{\dag}\ \Sigma\right)
\Tr \left(  \lambda^1_2 \Sigma^{\dag} \right)  \right]
 \label{chi-lag} \, ,
\eea
where $\lambda^i_j$ are combinations of Gell-Mann
$SU(3)$ matrices defined by $(\lambda^i_j)_{lk} = \delta_{il}\delta_{jk}$
and $\Sigma$ is defined in \eq{sigma}.
The covariant
derivatives in \eq{chi-lag} are taken with respect to the external
gauge fields whenever they are present.

The convenience of the non-standard form of \eq{chi-lag} will become
clear in the following.
In our approach the lagrangian (\ref{chi-lag}) is the effective theory
generated by integration of the three light quarks
of the $\Delta S=1$ quark lagrangian.
Therefore the  notation is such as to remind us the flavor and the
chiral structure of the quark operators.
In particular,
$G_{\underline{8}}$ represents
the $(\underline{8}_L \times \underline{1}_R)$ part of the interaction,
as it is induced in QCD by the
gluonic penguins, while the two terms
proportional to $G_{LL}^a$ and $G_{LL}^b$ are admixtures of
the $(\underline{27}_L \times \underline{1}_R)$ and the
$(\underline{8}_L \times \underline{1}_R)$ part of the interaction,
 such as it is induced by left-handed
 current-current operators; the term proportional to $G^{(0)}$ is
the constant (non-derivative) part arising in the isospin violating and
 $(\underline{8}_L \times \underline{8}_R)$ electroweak
components, of which the terms proportional to $G_{LR}^{a,b,c}$ 
represent the $O(p^2)$ momentum corrections.

The terms proportional to $G_{\underline{8}}$, $G^a_{LL}$ and $G^b_{LL}$ have
been already studied in the literature~\cite{Cronin,PdeR,EKW}
in the framework of chiral
perturbation theory. Those proportional to $G^a_{LL}$ and $G^b_{LL}$
are usually separated further into their isospin  components (see the appendix)
as
\bea
 {\cal L}_{\underline{27}} & = & g_{\underline{27}} \: \left[
 \frac{2}{3}\,  \Tr \left( \lambda^1_2 \Sigma^{\dag}
D_\mu  \Sigma \right)
\Tr \left(  \lambda^3_1 \Sigma^{\dag} D^\mu \Sigma \right) \right. \nnu \\
 & & + \left. \: \Tr \left( \lambda^3_2 \Sigma^{\dag} D_\mu
\Sigma \right)
\Tr \left(  \lambda^1_1 \Sigma^{\dag} D^\mu \Sigma \right) \right] \, ,
\label{a27}
\eea
which is a $(\underline{27}_L \times \underline{1}_R)$, and
\bea
 {\cal L}_{\underline{8}} & = &
g_{\underline{8}} \: \left[ \Tr \left( \lambda^1_2 \Sigma^{\dag}
D_\mu  \Sigma \right)
\Tr \left(  \lambda^3_1 \Sigma^{\dag} D^\mu \Sigma \right) \right. \nnu \\
 & & - \: \left. \Tr \left( \lambda^3_2 \Sigma^{\dag} D_\mu
\Sigma \right)
\Tr \left(  \lambda^1_1 \Sigma^{\dag} D^\mu \Sigma \right) \right] \, ,
\label{a8}
\eea
which is a pure $(\underline{8}_L \times \underline{1}_R)$.
We prefer to keep the  $\Delta S = 1$ chiral Lagrangian in the form given in
\eq{chi-lag}, which makes the bosonization of the various quark operators more
transparent, and
perform the isospin projections at the level of the matrix elements.
Equations (\ref{a27})--(\ref{a8}) provide anyhow the
comparison with other references. The chiral coefficients in the two approaches
are related by
\bea
g_{\underline{8}} (Q_{1,2}) & = & \frac{1}{5} \left[ 3 \: G^a_{LL}(Q_{1,2}) -
2 \:
G^b_{LL} (Q_{1,2}) \right] \nnu \\
g_{\underline{27}} (Q_{1,2}) & = & \frac{3}{5} \left[ G^a_{LL}(Q_{1,2}) +
G^b_{LL} (Q_{1,2}) \right] \, .
\eea

Concerning the $(\underline{8}_L \times \underline{8}_R)$ part, the constant 
term was 
first considered in~\cite{BW}; its momentum corrections are discussed here
for the first time, as far as we know.

In this paper we only need the  weak chiral lagrangian to $O(p^2)$. The
 $O (p^4 )$ weak lagrangian is very complicated with thirty-seven~\cite{EKW}
coefficients to be
determined only in the $(\underline{8}_L \times \underline{1}_R)$ sector, and
many more in the others.
Terms
proportional to the current quark masses
belong to such a higher-order lagrangian---except for a
term $O(m_q)$ for the operators $Q_{7,8}$:
\beq
G^{(m)} (Q_{7,8})\ \left[ \Tr \left( \lambda^3_2 \Sigma^\dag \lambda _1^1 
\Sigma {\cal M}^\dag \Sigma \right) + 
\Tr \left(\lambda_1^1 \Sigma  \lambda^3_2  \Sigma^\dag 
{\cal M} \Sigma^\dag \right)\right] \, ,
\eeq
whose contribution is small compared to the leading constant term.

The weight of some of the next-to-leading order corrections on the physical
amplitudes has been estimated in
various models~\cite{EKW,Fej} as well as in the
$\chi$QM~\cite{BEF}. They range from 10 to $30 \%$ of the leading
contributions, a potentially large
effect that we should bear in mind when estimating the inherent uncertainty
 of our computation.

\section{The Model at Work}

The seven coefficients $G_i$ in (\ref{chi-lag})
would have in general to be determined phenomenologically. In the $\chi$QM,
 they are calculable by integration of the constituent quarks.
This section contains those  details of the
computation that require some explicit discussion, since a few subtle points
arise in the regularization procedure.

\subsection{Regularization}

Some of the constituent-quark loops are divergent.
We use dimensional regularization (with the definition $d=4 - 2\epsilon$).
The logarithmic and quadratic divergencies
of the loop integration are identified with, respectively, the pion decay
constant $f_\pi$ and the quark condensate $\langle \bar{q} q \rangle$.
We define the two quantities
\bea
f^{(0)} &= &\frac{M^2 N_c}{4 \pi^2 f}
\left( \frac{4 \pi \tilde{\mu}^2}{M^2} \right) ^\epsilon \Gamma (\epsilon )
 \, , \label{fF} \\
\vev{\bar{q} q}^{(0)} & = & - \frac{M^3 N_c}{4 \pi^2}
\left( \frac{4 \pi \tilde{\mu}^2}{M^2} \right) ^\epsilon\Gamma ( -1 +
\epsilon )\ ,
\label{qqF}
\eea
and then identify $f^{(0)} = f_\pi$ and $\vev{\bar{q}q}^{(0)}
= \vev{\bar{q}q}$ in the end of the computation. The latter are understood as
the physical quantities, inclusive of gluon  and mass corrections. In
particular,
a direct computation within the model shows that
\beq
\langle \bar{q}q \rangle^{(\alpha_s)} =  \langle \bar{q}q \rangle ^{(0)} -
\frac{1}{12 M}
\langle \frac{\alpha_s}{\pi} GG \rangle\ \ +\ \
\mbox{(higher gluon condensates)} \, ,
\eeq
where we neglect current quark mass corrections. Similarly
for $f^{(0)}$.
We use \eqs{fF}{qqF} as a convenient
bookkeeping device for these input parameters.

There is a certain degree of arbitrariness in these definitions inasmuch as
the two gamma functions go into each other up to a
finite term by expanding around the pole.
Yet, for all practical purposes, keeping the two types of divergencies
separated is a simple and effective way of singling out the different
contributions.

Since the $\tilde{\mu}$ dependence
in eqs. (\ref{fF}) and (\ref{qqF}) is
absorbed in the physical quantities $f_\pi$ and
$\vev{\bar{q}q}$~\fnote{\dag}{The quark condensate has a
perturbative scale dependence
which originates in the short-distance computation; see the discussion of
section 7},
there is no scale dependence induced by the $\chi$QM in the 
chiral coefficients.

The $\chi$QM can be thought as an effective QCD model
interpolating between $\Lambda_\chi$ and $M$. 
In this sense, the most natural
 regularization scheme is a cut-off theory where 
no divergencies arise. In fact, one finds
\bea
f^{(0)} &= &\frac{M^2 N_c}{4 \pi^2 f}
\ln \frac{\Lambda_\chi^2}{M^2} \, ,\label{a} \\
\vev{\bar{q} q}^{(0)} & = & \frac{M N_c}{4 \pi^2}
\left( - \Lambda_\chi^2 + M^2 \ln \frac{\Lambda_\chi^2}{M^2} \right) \, .
\label{b}
\eea
Eqs.~(\ref{a}) and (\ref{b}) are finite (albeit cut-off dependent)
quantities to be identified with $f_\pi$ and $\vev{\bar{q} q}$, respectively.
Eq.~(\ref{b}) can be understood in the ENJL model
as a solution of the Schwinger-Dyson equation
 with gap $M$.
Notice
that in this case $M$ exhibits an intrinsic scale dependence
and it vanishes at energies higher than $\Lambda_\chi$.

Unfortunately, such a cut-off regularization of the hadronic
matrix elements is not consistent with that of the Wilson
coefficients, that are obtained by means of dimensional regularization. For
this reason,  we do not pursue this possibility further (see, however, refs.
\cite{BBdeR}, where a matching of cut-off regularization and NDR was
attempted).

The scale dependence---that
must appear in the hadronic matrix elements in order to make
physical quantities scale independent---is introduced in our approach
by the meson loop corrections to the hadronic matrix elements, that we
calculate in section 8 for the $K\to\pi\pi$ amplitudes.

\subsection{The Computation}

Within the model, any computation is easily
performed  by applying the Feynman rules given in the
appendix.

As shown in fig. 1,
for each operator there are two possible configurations that
must be estimated. Whether the unfactorized form of fig. 1(A) or the factorized
form of fig. 1(B) is the leading one in the $1/N_c$ expansion depends on the
color structure of
the operator. For example, the unfactorized configuration
is the leading one for the color ``unsaturated'' $Q_1$,
 but  the opposite is true
for the color ``saturated'' operator $Q_2$.
\begin{figure}[t]
\epsfxsize=10cm
\centerline{\epsfbox{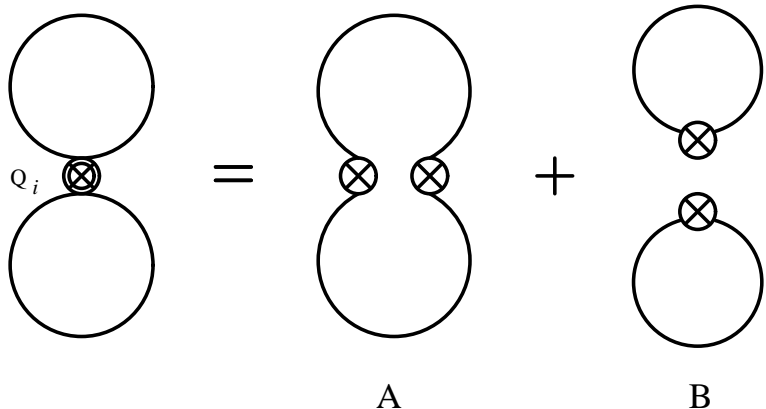}}
\caption{The quark constituent loops for an arbitrary
operator. (A) is the unfactorized pattern, (B) the factorized one. The crossed
circles represent operator and/or current insertions.}
\end{figure}

The coefficients $G^{a,b}_{LL}$, for example, are most easily computed
by analyzing the two-meson diagrams of fig. 2.
In the case of $Q_1$,
we may use the transition $K^0 \rightarrow \pi^0$ in order to fix
uniquely $G_{LL}^b (Q_1)$ and $K^+ \rightarrow \pi^+$ to fix
$G_{LL}^a (Q_1)$ (see the Feynman rules of table 1 in appendix B).
In the HV scheme and before gluon corrections
they turn out to be
\bea
G_{LL}^b (Q_1) & = & -f^2 (f^{(0)} )^2 \nnu \\
G_{LL}^a (Q_1) & = & -f^2 (f^{(0)} )^2 /N_c \, ,
\eea
which is the expected pattern in the $1/N_c$ expansion.

The diagrams with three meson external lines, which are the
relevant ones in the computation of the matrix elements of section 8, are then
generated by
means of the chiral lagrangian in \eq{chi-lag}.

\begin{figure}[t]
\epsfxsize=14cm
\centerline{\epsfbox{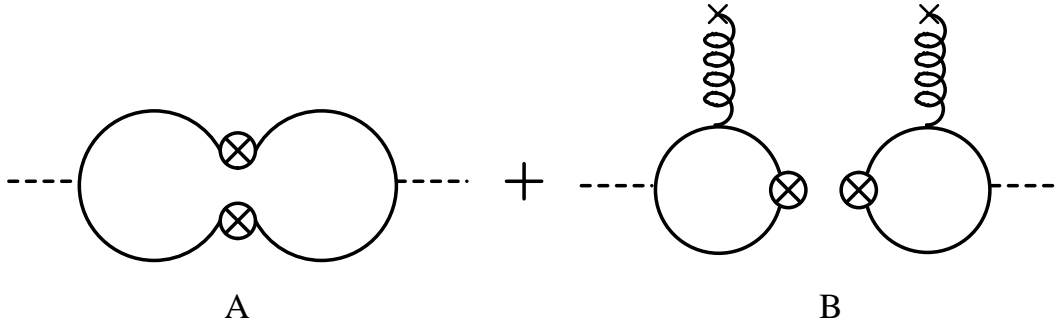}}
\caption{The constituent quark loops coupled to the mesons for an arbitrary
operator. (A) is the unfactorized pattern, (B) the factorized one with
the insertion of the gluon condensate (in this case only the color
octet components, proportional to $T^a$, of the currents contribute).
Meson and gluon lines are attached in all possible ways.}
\end{figure}

\subsection{Non-Perturbative Gluonic Corrections}

Non-perturbative gluonic corrections are  introduced by
propagating the quarks
in the background of external soft gluons, as suggested
in ref.~\cite{PdeR}. They provide a sizable correction of order $O(\alpha_s
N_c)$
that we parameterize in terms of the leading gluon condensate $\vev{\alpha_s
GG/\pi}$ by introducing the quantity
\beq
 \delta_{\langle GG \rangle} = \frac{N_c}{2} \frac{\langle
 \alpha_s G G/\pi \rangle}{16 \pi^2 f^4} \, . \label{GG}
\eeq

Since this correction is finite we can always compute it in the
factorized form by means of a Fierz
transformation (see \eqs{diracF}{colorF} in the appendix).
In this way it is easier to single out those contributions that are
non-vanishing in the presence of the external gluons. The
leading gluon condensate contribution is of $O(1/N_c)$ and it is generated
by diagrams of the type shown in fig. 2(B).

Only those configurations in which one external
gluon field is attached to each quark loop are genuine corrections to the
matrix element, since those in which both gluons are attached
to the same quark loop are included in the renormalization of either the quark
condensate or $f_\pi$, for which we take the physical values.

\subsection{$\gamma_5$-Scheme Dependence}

In dealing with
the problem of $\gamma_5$ in dimensional regularization,
we have con\-si\-dered both,
the NDR sche\-me---anti-commu\-ting $\gamma_5$
in $d$ dimensions---and the
HV scheme---commuting $\gamma_5$ in $d \neq 4$ dimensions. Both procedures
have been used to obtain a consistent set of NLO anomalous dimension
matrices
for the perturbative evolution of the Wilson coefficients~\cite{Monaco,Roma}.

Two questions related to the $\gamma_5$-scheme dependence
need addressing before proceeding with the actual computation. To these we now
turn.

\subsubsection{Fake Anomalies in the HV scheme}

A first problem arises in considering
the building blocks out of which the matrix elements
in the factorized form of fig. 1(B) are made.
These have an independent physical interpretation, as we shall see, and come in
four kinds. Let us consider first the result of the NDR scheme. There are two
densities, that we need to $O (p^2)$, for which we obtain
 \bea
 \vev{ 0|\,\overline{s} \gamma_5 u\,
|K^+(k)}_{\mbox{\scriptsize NDR}} &  =&  i \sqrt{2} \;
\left[ \frac{\vev{\overline{q} q}^{(0)}}{f} - k^2 \,
\frac{f^{(0)}}{2 M} \right]
\, , \label{Kvacuum-NDR} \\
\langle \pi^+(p_+)|\,\overline{s} d\, |K^+(k)
\rangle_{\mbox{\scriptsize NDR}} \, &=&
\, - \frac{\langle \overline{q} q
\rangle ^{(0)}}{f^2}
 +  \frac{3 M}{2 \Lambda_\chi^2} P^2 \nnu \\
 & & +
\, \frac{q^2}{2 M} \left( f_+^{\mbox{\scriptsize NDR}}(0) \,
\, - 3\, \frac{M^2}{\Lambda_\chi^2} \right)
\, ,     \label{Kpi-NDR}
\eea
and two currents, that we only need to $O(p)$:
\bea
\vev{ 0|\,\overline{s} \gamma^\mu \gamma_5 u\,
|K^+(k)}_{\mbox{\scriptsize NDR}} &  =&  i \sqrt{2} \;
k^\mu f^{(0)}
\label{fpiNDR} \\
 \langle \pi^+(p_+)|\,\overline{s}\gamma ^\mu d\, |K^+(k)
\rangle_{\mbox{\scriptsize NDR}} \, &=& -
f_+^{\mbox{\scriptsize NDR}}(q^2)P^\mu + f_-(q^2)q^\mu \label{f-NDR} \, ,
\label{f+NDR}\,
\eea
where  $q=k-p_+$ and $P=k+p_+$,
while $f_+^{\mbox{\scriptsize NDR}}(0) \equiv f^{(0)}/f $ is
 identified with the vector form factor at
zero momentum transfer $q$.

In the NDR scheme we correctly find
\beq
f_+^{\mbox{\scriptsize NDR}} (0) = 1 \qquad \mbox{and} \qquad f_-(0) = 0 \, .
\label{f+} \eeq
This result holds in the limit of unbroken $SU(3)$ symmetry, where
\beq
f_\pi = f_K \qquad \mbox{and} \qquad f^{K^0\pi^0}_\pm (0) =
f^{K^+\pi^0}_\pm (0)
= f_\pm (0) \, .
\eeq
In this limit, eqs.~(\ref{f+}) are in agreement with the experiments
(deviations
of $f_+(0)$ from unity are of order $m_s^2$~\cite{AG}) and
 we do not find any anomalous result in the NDR scheme.

As we consider next the HV scheme,
the two densities are now given by
 \bea
 \vev{ 0|\,\overline{s} \gamma_5 u\, |K^+(k)}_{\mbox{\scriptsize HV}} &  =&
  \vev{ 0|\,\overline{s} \gamma_5 u\, |K^+(k)}_{\mbox{\scriptsize NDR}}
\nnu \\
& &  + \: i \:\sqrt{2}\ f\
\left[
12 \, \frac{M^3}{\Lambda_\chi^2}\left(1 - \frac{k^2}{6 M^2}\right) \right]
\, ,
\label{Kvacuum-HV} \\
 \vev{ \pi^+ (p_+)|\,\overline{s} d\, |K^+(k)}_{\mbox{\scriptsize HV}} &  =&
  \vev{ \pi^+ (p_+)|\,\overline{s} d\, |K^+(k)}_{\mbox{\scriptsize NDR}}
 +  24\,
\frac{M^3}{\Lambda_\chi^2}
\, . \label{Kpi-HV}
\eea
while for the current matrix elements we find
\bea
\vev{ 0|\,\overline{s} \gamma^\mu \gamma_5 u\,
|K^+(k)}_{\mbox{\scriptsize HV}} &  =&  i \sqrt{2} \;
k^\mu f^{(0)}
\label{fpiHV} \\
 \langle \pi^+(p_+)|\,\overline{s}\gamma ^\mu d\, |K^+(k)
\rangle_{\mbox{\scriptsize HV}} \, &=& -
f_+^{\mbox{\scriptsize HV}}(q^2)P^\mu + f_-(q^2)q^\mu \, ,
\label{f+HV}\,
\eea
where
 \beq
f_+^{\mbox{\scriptsize HV}}(0) = 1 + 4 \,\frac{M^2}{\Lambda_\chi^2}\, ,
\label{fplusHV}
\eeq

Eq.~(\ref{fpiHV}) is the same as \eq{fpiNDR},
so that $f_\pi$ is defined identically in the two renormalization
schemes. On the other hand, the vector form factor $f_+(0)$ 
turns out to be different,
as it happens for the density matrix elements.

Even though $f_+^{\mbox{\scriptsize HV}}(0) \neq
1$ the vector current Ward identity is preserved also in the HV scheme.
In fact, in the model
the mesons propagate via quark loops, so that a simple
calculation in the NDR case leads to
\beq
 G^{-1}_\Pi (k) = k^2 f_+^{\mbox{\scriptsize NDR}}(0)
\label{prop-NDR}\, ;
\eeq
while in the HV we find
\beq
 G^{-1}_\Pi (k) = k^2 f_+^{\mbox{\scriptsize HV}}(0)
                  - 24\ \frac{M^2}{\Lambda_\chi^2}\ M^2  \, .
\label{prop-HV}
\eeq
By looking at the term proportional to $k^2$ we see by inspection that the
same shift in $f_+(0)$
is present in the propagator as well as in the vertex.
Therefore the vector Ward identity 
\beq
q^\mu \langle \pi^+(p_2)|
\frac{2}{3}{\bar{u}\gamma_\mu u} 
-\frac{1}{3}{\bar{d}\gamma_\mu d} 
|\pi^+(p_1) \rangle =
      G^{-1}_\pi (p_2) - G^{-1}_\pi  (p_1)
\label{WI}
\eeq
holds in both schemes,
and one might think that a redefinition
of $f_+^{\mbox{\scriptsize HV}}(0)$
could solve all problems. Unfortunately, because $f^{(0)}$ is
the same in the two schemes, it is not possible to simply redefine
$f_+^{\mbox{\scriptsize HV}}(0)$ to be equal to~1, reabsorbing the
HV shift in $f_\pi$.
Moreover, \eq{prop-HV} shows another problem:
a mass term is
generated in the HV case, thus leading to  explicit
breaking of chiral symmetry~\fnote{\dag}{There has
been  some discussion in the literature
about similar problems in the standard model. For instance, in ref.~\cite{HV1}
chiral anomalies in addition to the ABJ's
were noticed. In ref.~\cite{HV2}, the need of subtractions was pointed out and
some
claims of ref. \cite{HV1} adjusted.}.

We therefore find that, in order to maintain the symmetries of the theory, we
must subtract appropriate
terms
in the HV case. In particular, the mass term  in \eq{prop-HV} must be
taken away. This subtraction leads to an analogous subtraction
in the building blocks of \eqs{Kvacuum-HV}{Kpi-HV}, which become identical to
the NDR results.
Having fully subtracted the propagators,
the Ward identity in (\ref{WI}) implies that
also $f_+^{\mbox{\scriptsize HV}}(0)$ in \eq{f+HV}
becomes identical to the NDR result.

The overall effect of the subtractions is to make the building blocks of the HV
scheme identical to those of NDR, as given in \eqs{Kvacuum-NDR}{f-NDR}.

These subtractions can be implemented in the strong sector
by the addition of appropriate terms in the $\chi$QM one-loop
chiral lagrangian that becomes, in the HV case:
\beq
{\cal L}_{\rm strong}^{(2)}
- \frac{f^2}{4} \, a_1 \, \Tr \left[ D^\mu \Sigma  D_\mu \Sigma^{\dag}\right]
- \frac{f^2}{4} \, a_2 \, \Tr \left[ \Sigma^{\dag} + \Sigma \right]
\label{chi-HV}
\eeq
where
\bea
a_1 & = & f_+^{\mbox{\scriptsize HV}}(0) - f_+^{\mbox{\scriptsize NDR}}(0) \\
a_2 & = &  \ 24\ \frac{M^2}{\Lambda_\chi^2}\ M^2 \, .
\eea
The lagrangian (\ref{chi-HV}) suffices in giving the correct propagator to
one loop. Notice that the term proportional to $a_2$ breaks explicitly chiral
invariance.

Other terms, containing the appropriate flavor projectors,
 must be added in order to eliminate  fakes anomalies
arising in the weak sector.

\subsubsection{Fierz Transformation in the two schemes}

Another related problem is the following.
By a Fierz transformation it is
possible to bring all matrix elements (leading as well as next-to-leading
 order in $1/N_c$) to
a factorized form.
As shown in the previous subsection,
the factorized building blocks (after subtraction) are scheme independent.
Therefore it would seem that there is no $\gamma_5$-scheme dependence at all
in the hadronic matrix elements.
Yet, this conclusion is not correct because,
 while in the HV scheme it is possible to apply a Fierz
transformation on the quark operators without changing the matrix element,
this is not possible in the NDR scheme where operators
related by a Fierz transformation lead to 
different matrix elements. This fact is at the origin of
the $\gamma_5$-scheme dependence of the matrix elements.

\begin{figure}[t]
\epsfxsize=14cm
\centerline{\epsfbox{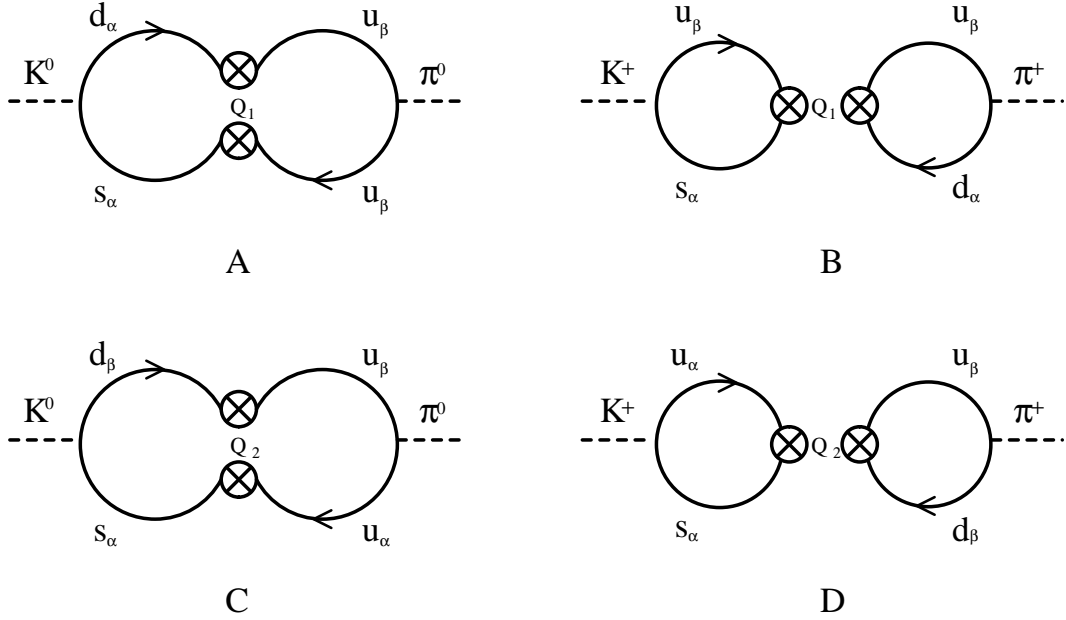}}
\caption{The constituent quark loops coupled to $K$ and $\pi$ with the
insertion of the operators $Q_1$ or $Q_2$, neglecting soft gluon corrections.
(A) gives the leading $O(N_c^2)$ contribution of $Q_1$ to the chiral
coefficient $G^b_{LL}$, while (B) is the subleading $O(N_c)$ correction
to $G^a_{LL}$.
The opposite happens for $Q_2$ due to the different color structure:
(C) gives the subleading contribution to $G^b_{LL}$ and (D)
 the leading one to $G^a_{LL}$. }
\end{figure}
As a first example, let us consider the contribution of $Q_1$ to
$G^b_{LL}$. A direct evaluation of the unfactorized
diagram in fig. 3(A) gives, in the NDR scheme,
\beq
 G^b_{LL} (Q_1) = - f^2 (f^{(0)} )^2 \left( 1 - 6\,
\frac{M^2}{\Lambda_\chi^2}
 \right) \, . \label{shifto}
 \eeq
 In the HV case we obtain instead
 \beq
  G^b_{LL}(Q_1) =  - f^2 (f^{(0)} )^2 \label{c827HV}
  \eeq

  We could have performed the same computation
 by first Fierz transforming the
operator $Q_1$ that becomes
  \beq
  \widetilde{Q}_1 = (\bar{s} d)_{V-A} (\bar{u}u)_{V-A}.
  \eeq
 In this case, by means of the 
building blocks in \eqs{fpiNDR}{f+NDR} (or the subtracted \eqs{fpiHV}{f+HV}),
 we would have obtained the result in
\eq{c827HV} in both the NDR
and HV schemes. We explicitly see that the Fierz related
operators $Q_1$ and $\widetilde{Q}_1$ might
lead in the NDR  scheme
to different contributions to the chiral coefficients.

As a second, more complicated example, let us take the contribution
of the operator $Q_6$ to $G_{\underline 8}$ and evaluate its bosonization
directly in the form given in \eq{Q1-10}. By computing
the two-loop unfactorized diagram in the NDR scheme we obtain
\beq
G_{\underline{8}} (Q_6) = 2 \frac{\vev{\bar{q}q}}{M} f^2 \left(
f_+^{\mbox{\scriptsize NDR}} - 9\,
\frac{M^2}{\Lambda_\chi^2}
\right)
\eeq
By performing a Fierz transformation on the operator
\beq
\widetilde{Q}_6 = - 8 (\bar{s}_L q_R )( \bar{q}_R d_L )
\eeq
and using the density building blocks in \eqs{Kvacuum-NDR}{f+NDR}
one obtains, in the NDR case,
\beq
G_{\underline{8}} (\widetilde{Q}_6) =
2 \frac{\vev{\bar{q}q}}{M} f^2 \left( f_+^{\mbox{\scriptsize NDR}} - 6\,
\frac{M^2}{\Lambda_\chi^2}
\right)\ .
\eeq
In the HV case instead both procedures lead to
\beq
G_{\underline{8}} (Q_6) = 2 \frac{\vev{\bar{q}q}}{M} f^2 \left(
f_+^{\mbox{\scriptsize HV}} + 6\,
\frac{M^2}{\Lambda_\chi^2}
\right)\ .
\label{c8HV}
\eeq

Notice that \eq{c8HV} is not the correct result because of the necessary
subtractions
that must still be implemented in the HV scheme, as discussed in the previous
subsection.

Because a similar pattern holds for all ten operators, we have proved that
a Fierz transformation preserves the $\chi$QM result in the HV scheme but not
 in the NDR one.
This feature, which was already observed in short-distance calculations
\cite{Monaco,Roma},
can be understood as a consequence of the prescription of symmetrization
of the chiral vertices in the HV scheme, which is equivalent to considering
the inserted operators as four dimensional objects, for which the usual Fierz
transformation are allowed.

As a consequence of these results, when computing matrix elements in the HV
scheme we will always resort to Fierzing the quark operators in such a way
to exploit the simpler factorized form. Subtractions are then applied
in order to satisfy the relevant Ward identities and equations of motion.
On the contrary, we do not apply any Fierz transformation in the NDR case.
We retain only terms
up to order $M^2/\Lambda_{\chi}^2$. Ambiguities in the
subtraction procedure---like those arising from the expansion
of the gamma functions in the building blocks
of fig. 1(B)---can be shown to be of higher order.

A subtlety  arises at this point.
The NDR prescription of ref.~\cite{Monaco,Roma}
 preserves the chiral properties of the 
operator ($Q_1 - Q_2$) by means of 
a special choice of  coefficients for the evanescent operators. 
Consistency with such a prescription suggests that 
we have to impose, by an appropriated 
subtraction, that the operator
$Q_1 - Q_2$ (as well as $Q_9 - Q_{10}$)
remains a pure octet $(\underline{8}_L \times \underline{1}_R$). 
As a consequence, the shift in \eq{shifto} is cancelled and
the matrix elements induced
by $Q_1$ and $Q_2$  are the same
in the two schemes. 

\section{The Coefficients of the Weak Chiral Lagrangian}

The results of the previous sections makes it possible to compute the
contribution
of each of the ten operators in \eq{Q1-10} to  the
seven  coefficients of the weak chiral lagrangian (\ref{chi-lag}) in both the
HV and the NDR
schemes.
We have included in our computation all contributions of
order $O (N_c^2)$, $O (N_c)$ and $O (\alpha_s N_c)$.
Our result  depends on the  intrinsic $\chi$QM
parameter $M$ and three input parameters: $f_{\pi}$, $\vev{GG}$
and $\vev{\bar{q}q}$.

For the purpose of this computation,
it is convenient to rewrite the electroweak operators as follows
\bea
Q_7 & = & \frac{3}{2} \hat{e}_d Q_5 + \frac{3}{2} ( \hat{e}_u -
\hat{e}_d ) \Delta Q_7 \nnu \\
Q_8 & = & \frac{3}{2} \hat{e}_d Q_6 + \frac{3}{2} ( \hat{e}_u -
\hat{e}_d ) \Delta Q_8 \nnu \\
Q_9 & = & \frac{3}{2} \hat{e}_d Q_3 + \frac{3}{2} ( \hat{e}_u -
\hat{e}_d ) \Delta Q_9 \nnu \\
Q_{10} & = & \frac{3}{2} \hat{e}_d Q_4 + \frac{3}{2} ( \hat{e}_u -
\hat{e}_d ) \Delta Q_{10} \, ,
\label{deltap1}
\eea
where
\bea
\Delta Q_7 & = & \left( \overline{s} d  \right)_{\rm V-A}
            \left( \overline{u}  u \right)_{\rm V+A} \nnu \\
\Delta Q_8 & = & \left( \overline{s}_{\alpha} d_{\beta}  \right)_{\rm V-A}
            \left( \overline{u}_{\beta}  u_{\alpha} \right)_{\rm V+A} \nnu
\\
\Delta Q_9 & = & \left( \overline{s} d  \right)_{\rm V-A}
            \left( \overline{u}  u \right)_{\rm V-A} \nnu \\
\Delta Q_{10} & = & \left( \overline{s}_{\alpha} d_{\beta}  \right)_{\rm V-A}
            \left( \overline{u}_{\beta}  u_{\alpha} \right)_{\rm V-A} \, .
\label{deltap2}
\eea
{}From now on, we identify $f^{(0)} = f = f_\pi$.
The inclusion in the chiral coefficients of
the corresponding factors
\beq
- \frac{G_F}{\sqrt{2}} V_{ud} V_{us}^{*}  
\Bigl[z_i(\mu) + \tau y_i(\mu) \Bigr]
\eeq
is understood.

\subsection{The HV result}

Let us first list the results for the chiral coefficients in the HV scheme.
We include in the coefficient $G_{\underline{8}}$ the contributions of
the gluonic penguins, which are pure octets, as well as the
gluon-penguin-like components of the electroweak penguins (\eq{deltap1}).
We therefore find:
\bea
G_{\underline{8}} (Q_3) & = &   f_\pi^4 \frac{1}{N_c} \left( 1 -
\delta_{\vev{GG}}
\right) \nnu \\
G_{\underline{8}} (Q_4) & = &   f_\pi^4 \nnu \\
G_{\underline{8}} (Q_5) & = &  \frac{2}{N_c} \,
\frac{\vev{\bar{q}q}}{M} f_\pi^2 \, \left( 1 - 6\,
\frac{M^2}{\Lambda_{\chi}^2} \right) \nnu \\
G_{\underline{8}} (Q_6) & = & 2 \,
\frac{\vev{\bar{q}q}}{M} f_\pi^2 \, \left( 1 - 6\,
\frac{M^2}{\Lambda_{\chi}^2} \right)  \nnu \\
G_{\underline{8}} (Q_7) & = &    3\, \hat{e}_d \frac{1}{N_c} \,
\frac{\vev{\bar{q}q}}{M} f_\pi^2 \, \left( 1 - 6\,
\frac{M^2}{\Lambda_{\chi}^2} \right)\nnu \\
G_{\underline{8}} (Q_8) & = &    3\, \hat{e}_d \,
\frac{\vev{\bar{q}q}}{M} f_\pi^2 \, \left( 1 - 6\,
\frac{M^2}{\Lambda_{\chi}^2} \right) \nnu \\
G_{\underline{8}} (Q_9) & = &   \frac{3}{2} \hat{e}_d \frac{1}{N_c}f_\pi^4
\left( 1
-\delta_{\vev{GG}} \right) \nnu \\
G_{\underline{8}} (Q_{10}) & = &   \frac{3}{2} \hat{e}_d f_\pi^4  \, ,
\eea
where $\delta_{\vev{GG}}$ is given by \eq{GG}.

The $(V-A)\otimes (V-A)$ operators $Q_{1,2}$ and $\Delta Q_{9,10}$ yield:
\bea
G_{LL}^a(Q_1) & = & - \frac{1}{N_c} f_\pi^4  \left( 1 - \delta_{\vev{GG}}
\right) \nnu \\
G_{LL}^a(Q_2) & = & - f_\pi^4  \nnu \\
G_{LL}^b(Q_1) & = & - f_\pi^4  \nnu \\
G_{LL}^b(Q_2) & = & - \frac{1}{N_c} f_\pi^4  \left( 1 - \delta_{\vev{GG}}
\right) \nnu \\
G_{LL}^a(Q_9) & = & - \frac{3}{2} ( \hat{e}_u - \hat{e}_d) f_\pi^4
\frac{1}{N_c}
 \left( 1 - \delta_{\vev{GG}} \right)\nnu \\
G_{LL}^a(Q_{10}) & = & - \frac{3}{2} ( \hat{e}_u - \hat{e}_d) f_\pi^4  \nnu \\
G_{LL}^b(Q_9) & = & - \frac{3}{2} ( \hat{e}_u - \hat{e}_d) f_\pi^4 \nnu \\
G_{LL}^b(Q_{10}) & = & - \frac{3}{2} ( \hat{e}_u - \hat{e}_d) f_\pi^4
\frac{1}{N_c}
 \left( 1 - \delta_{\vev{GG}} \right) \, .
 \eea

 For the constant term which represent the leading contribution
of $Q_8$ and $Q_7$, we find:
 \bea
 G^{(0)}(Q_7) & = &  - 3
 \frac{1}{N_c} ( \hat{e}_u - \hat{e}_d) \vev{\bar{q}q}^2  \nnu \\
 G^{(0)}(Q_8) & = &  - 3  ( \hat{e}_u - \hat{e}_d) \vev{\bar{q}q}^2
\eea
while, for their momentum corrections, we have:
\bea
 G_{LR}^a(Q_7) & = & 3  ( \hat{e}_u - \hat{e}_d) \frac{1}{N_c}
 \frac{f_\pi^2}{M}
 \vev{\bar{q}q} \nnu \\
G_{LR}^a(Q_8) & = &  3  ( \hat{e}_u - \hat{e}_d) \frac{f_\pi^2}{M}
 \vev{\bar{q}q} \nnu \\
 G_{LR}^b(Q_7) & = &  - \frac{3}{2}  ( \hat{e}_u - \hat{e}_d)
 f_\pi^4 \nnu \\
 G_{LR}^b(Q_8) & = &  - \frac{3}{2}   ( \hat{e}_u - \hat{e}_d)
 f_\pi^4  \frac{1}{N_c} \left(1 + \delta_{\vev{GG}}\right) \nnu \\
  G_{LR}^c (Q_7) & = & - 9\ \frac{M^2}{\Lambda^2_\chi}
  ( \hat{e}_u - \hat{e}_d) \frac{1}{N_c}
 \frac{f_\pi^2}{M}
 \vev{\bar{q}q} 
\nnu \\
G_{LR}^c (Q_8) & = & - 9\ \frac{M^2}{\Lambda^2_\chi}
  ( \hat{e}_u - \hat{e}_d) \frac{f_\pi^2}{M}
 \vev{\bar{q}q} \label{gcHV}
\, .
\eea

\subsection{The NDR result}

A similar computation in the NDR scheme leads to a different determination
of the coefficients because of the shifts discussed in the previous section. In
general, a shift is present whenever the determination of the coefficient
requires the evaluation of the unfactorized configuration of fig. 1(A).

The $G_{\underline{8}}$ contributions are now given by:
\bea
G_{\underline{8}} (Q_3) & = &  f_\pi^4 \frac{1}{N_c} \left( 1 -
\delta_{\vev{GG}}
- 6 \frac{M^2}{\Lambda_\chi^2}
\right) \nnu \\
G_{\underline{8}} (Q_4) & = &  f_\pi^4 \left( 1 - 6 \frac{M^2}{\Lambda_\chi^2}
\right) \nnu \\
G_{\underline{8}} (Q_5) & = &   \frac{2}{N_c} \,
\frac{\vev{\bar{q}q}}{M} f_\pi^2 \, \left( 1 - 9\,
\frac{M^2}{\Lambda_{\chi}^2} \right)\nnu \\
G_{\underline{8}} (Q_6) & = & 2 \,
\frac{\vev{\bar{q}q}}{M} f_\pi^2 \, \left( 1 - 9\,
\frac{M^2}{\Lambda_{\chi}^2} \right)  \nnu \\
G_{\underline{8}} (Q_7) & = &   3 \, \hat{e}_d \frac{1}{N_c}\,
\frac{\vev{\bar{q}q}}{M} f_\pi^2 \, \left( 1 - 9\,
\frac{M^2}{\Lambda_{\chi}^2} \right)\nnu \\
G_{\underline{8}} (Q_8) & = &   3 \, \hat{e}_d  \,
\frac{\vev{\bar{q}q}}{M} f_\pi^2 \, \left( 1 - 9\,
\frac{M^2}{\Lambda_{\chi}^2} \right) \nnu \\
G_{\underline{8}} (Q_9) & = &  \frac{3}{2} \hat{e}_d f_\pi^4 \frac{1}{N_c}
\left( 1 -\delta_{\vev{GG}} - 6 \frac{M^2}{\Lambda_\chi^2} \right) \nnu \\
G_{\underline{8}} (Q_{10}) & = &  \frac{3}{2} \hat{e}_d f_\pi^4 \left( 1
- 6 \frac{M^2}{\Lambda_\chi^2} \right) \, .
\eea

For the $(V-A)\otimes (V-A)$ operators we have:
\bea
G_{LL}^a(Q_1) & = & - \frac{1}{N_c} f_\pi^4  \left( 1 - \delta_{\vev{GG}}
\right) \nnu \\
G_{LL}^a(Q_2) & = & - f_\pi^4
\nnu \\
G_{LL}^b(Q_1) & = & - f_\pi^4 
\nnu \\
G_{LL}^b(Q_2) & = & - \frac{1}{N_c} f_\pi^4  \left( 1 - \delta_{\vev{GG}}
 \right) \nnu \\
G_{LL}^a(Q_9) & = & - \frac{3}{2} ( \hat{e}_u - \hat{e}_d) f_\pi^4
\frac{1}{N_c}
 \left( 1 - \delta_{\vev{GG}} \right)\nnu \\
G_{LL}^a(Q_{10}) & = & - \frac{3}{2} ( \hat{e}_u - \hat{e}_d) f_\pi^4 \nnu \\
G_{LL}^b(Q_9) & = & - \frac{3}{2} ( \hat{e}_u - \hat{e}_d) f_\pi^4 \nnu \\
G_{LL}^b(Q_{10}) & = & - \frac{3}{2} ( \hat{e}_u - \hat{e}_d) f_\pi^4
\frac{1}{N_c}
 \left( 1 - \delta_{\vev{GG}} \right)
 \eea
For the constant term we find:
 \bea
 G^{(0)}(Q_8) & = & - 3  ( \hat{e}_u - \hat{e}_d) \vev{\bar{q}q}^2
 \left( 1 - 3 \frac{M^3 f_\pi^2}{\vev{\bar{q}q} \Lambda_\chi^2} \right) \nnu \\
 G^{(0)}(Q_7) & = & - 3
 \frac{1}{N_c} ( \hat{e}_u - \hat{e}_d) \vev{\bar{q}q}^2
\left( 1 - 3 \frac{M^3 f_\pi^2}{\vev{\bar{q}q} \Lambda_\chi^2} \right)
\eea
with the corresponding momentum corrections:
\bea
 G_{LR}^a(Q_7) & = & 3 ( \hat{e}_u - \hat{e}_d) \frac{1}{N_c} 
\frac{f_\pi^2}{M}  \vev{\bar{q}q} \left( 1
- 3 \frac{M^2}{\Lambda_\chi^2} \right) \nnu \\
G_{LR}^a(Q_8) & = & 3  ( \hat{e}_u - \hat{e}_d) \frac{f_\pi^2}{M}
 \vev{\bar{q}q} \left( 1
- 3 \frac{M^2}{\Lambda_\chi^2} \right) \nnu \\
 G_{LR}^b(Q_7) & = &  -\frac{3}{2}  ( \hat{e}_u - \hat{e}_d)
 f_\pi^4 \nnu \\
 G_{LR}^b(Q_8) & = &  -\frac{3}{2}  ( \hat{e}_u - \hat{e}_d)
 f_\pi^4 \frac{1}{N_c}  \left(1 + \delta_{\vev{GG}}\right) \nnu \\
G_{LR}^c(Q_7) & = &  - 9 \frac{M^2}{\Lambda^2_\chi}
( \hat{e}_u - \hat{e}_d) \frac{1}{N_c} \frac{f_\pi^2}{M}
 \vev{\bar{q}q}  \nnu \\
G_{LR}^c(Q_8) & = &  - 9 \frac{M^2}{\Lambda^2_\chi}
  ( \hat{e}_u - \hat{e}_d) \frac{f_\pi^2}{M}
 \vev{\bar{q}q}  \, . \label{gcNDR}
 \eea

\subsection{Discussion}

It is perhaps useful  to summarize here the approximations we have made in
obtaining the
weak chiral lagrangian of \eq{chi-lag} and its
coefficients.

First of all, the lagrangian (\ref{chi-lag}) is only given to the
$O(p^2)$ and
any amplitude is expected to receive sizable corrections from higher order
terms.
These have been classified~\cite{EKW} but a direct computation of their
coefficients, even in such a simple model as the  $\chi$QM,
is a rather formidable task. An estimate of the $O(p^4)$ contributions
to some matrix elements of $Q_6$ has been given in ref. \cite{BEF}. The
NLO corrections vary from 10\% to 30\% depending on $M$.

Such an approximation implies  that we are also neglecting the
effect of the heavier meson multiplets.
As discussed in section 4,
they give
in principle rather large contributions in the strong sector, in particular
to $L_5$. In the
weak sector,
their effect seems to be less dramatic (see below)---except in the
modification they induce in the axial coupling $g_A$ between the constituent
quarks and the mesons. This is
clearly a place where future improvement is needed.

Another approximation we have made is in keeping only the first term in the
expansion in
$M^2/\Lambda^2_\chi$, that is characteristic of the model.
$M^2/\Lambda^2_\chi$ is small enough to make us confident of our result.
Anyway, insofar as the
$\chi$QM is just a (very) simple model, it is not clear whether going to the
next order would result in a real improvement.

Finally, the chiral coefficients are given to $O(1/N_c)$. In this respect
it is worth noticing that the quark operators
$Q_4$ and $Q_6$ do not induce any
$O(1/N_c)$ correction to the coefficients. This happens because of kinematics
and of the flavor singlet structure of the currents
(which induces cancellations among $u$ and $d$ flavor exchange in the
subleading configurations). As a consequence, gluonic corrections as well
are absent for $Q_4$ and $Q_6$, appearing first at $O(1/N_c^2)$.
Similarly, no 
gluonic corrections appear for the operators $Q_5$ and $Q_7$ at $O(1/N_c)$
because of their color and chiral structures. Notice that gluon
corrections are in the form $(1-\delta_{\vev{GG}})$ for $LL$ operators
and $(1+\delta_{\vev{GG}})$ for those with $LR$ current structure.

\section{The input parameters and M}

The quark and the gluon condensates are two  input parameters of
 our computation which are in principle free. 
 Their phenomenological determination  is a complicated
issue (they parameterize the genuine non-perturbative part of the
computation) and the literature offers different evaluations.

For guidance,
we identify the condensates entering our computation with those obtained by
fitting the experimental data  by means of the QCD sum rules 
(QCD-SR)~\cite{sumrules} or
lattice computations.
In our approach, we vary
these input parameters within the given ranges in order to
obtain a value for the amplitude we are interested in.

\subsection{Gluon Condensate}

For the gluon condensate, the most
 recent QCD-SR
analysis~\cite{Narison}, based on  $e^+e^-$ data, gives the
 scale independent result
\beq
\langle \frac{\alpha_s}{\pi} G G \rangle = (388 \pm  10 \:
\mbox{MeV} )^4 \, . 
\eeq
Such a value is consistent with older QCD-SR determinations \cite{oldGG}
as well as  with another recent one that finds~\cite{BNP}
 \beq
\langle \frac{\alpha_s}{\pi} G G \rangle = (376 \pm  47 \:
\mbox{MeV} )^4 \, . \label{GGEXP}
\eeq

These values are systematically smaller than the central value of the
lattice result~\cite{lattice}
\beq
\langle \frac{\alpha_s}{\pi} G G \rangle = (460 \pm  21 \:
\mbox{MeV} )^4 
\eeq 
which however suffers of a systematic error that is difficult
to evaluate. 
We will
take (\ref{GGEXP}) as the range to be explored in  numerical estimates.

\subsection{Quark Condensate}

The quark condensate is an important parameter in our computation because
it controls the size of the  penguin contributions, in particular of the
leading operator $Q_6$.
Unfortunately, 
the  uncertainty about its value  is  large
because of the sizable discrepancies among different estimates.

In the QCD-SR approach,  the quark condensate can
be determined  from $\Psi_5(q)$, the two-point function
 of the hadronic axial current,  as~\cite{DdeR}
\bea
\Psi_5 (0) & = & - (m_u + m_d ) \langle \bar{u}u + \bar{d}d \rangle \equiv
2 f_\pi^2 m_\pi^2 ( 1 - \delta_\pi ) \nnu \\
& = & ( 3.2 - 3.3 )  \times 10^8 \:
\mbox{MeV}^4 \, . \label{6.3}
\eea
This estimate agrees with the most recent determination~\cite{BPdeR} of the
parameter $\delta_\pi$ that quantifies deviations from the PCAC result. Such
deviations are of a few percents in \eq{6.3} (but larger in the case of
the strange-quark condensate).

These are scale independent values. To obtain the scale-dependent
quark condensate, we use the
renormalization-group  running masses $\overline{m}_u + \overline{m}_d$, the
value of
which has been estimated at 1 GeV to be~\cite{BPdeR}
\beq
\overline{m}_u + \overline{m}_d = 12 \pm 2.5 \: \mbox{MeV} \label{mumd}
\eeq
for $\Lambda^{(3)}_{\rm QCD} = 300 \pm 150$ MeV. The error
in
(\ref{mumd}) reflects changes in the spectral functions. 
In our numerical estimates,  we will take as input values the running masses at
1 GeV given by (\ref{mumd}). Even though our preferred range of
$\Lambda^{(4)}_{\rm QCD}$ (see \eq{lambdone}) corresponds to
$\Lambda^{(3)}_{\rm QCD} = 400 \pm 100 \: \mbox{MeV}$, we feel that we are
not making too large an error since this determination is not very
sensitive to the choice of $\Lambda_{\rm QCD}$.

By taking the value (\ref{mumd}),
we find  for the scale-dependent (and normal-ordered)
condensate
\beq
\vev{\bar{q}q} (\mu) = - \frac{f_\pi^2 m_\pi^2 ( 1 - \delta_\pi
)}{\overline{m}_u(\mu) + \overline{m}_d (\mu)} \, ,
\eeq\
 the numerical values of
\beq
\vev{\bar{q}q} = - (238 \pm 19 \: \mbox{MeV} )^3 
\label{qq1}
\eeq
at 1 GeV and
\beq
\vev{\bar{q}q} = - (222 \pm 19 \: \mbox{MeV} )^3
\label{qq2}
\eeq
at 0.8 GeV.
The error in \eqs{qq1}{qq2} is due to that in (\ref{mumd}). 

On the other hand, a recent
 determination of the quark condensate in 
lattice simulations  
 with quenched Wilson fermions~\cite{qqlattice1}
finds a value of
\beq
\vev{\bar{q}q} = - (257 \pm 27 \: \mbox{MeV} )^3
\label{qq3}
\eeq 
at 1 GeV. 

A similar simulation
with dynamical staggered fermions~\cite{qqlattice2}  yields the 
rather large result
 \beq
\vev{\bar{q}q} = - (380 \pm 7 \: \mbox{MeV} )^3
\label{qq4} \, ,
\eeq 
which is probably an overestimate to be discarded for our purposes.

As we can see, the actual  for the quark
condensate should be varied in the range 
\beq
 - ( 200 \: \mbox{MeV} )^3 \leq 
 \vev{\bar{q}q} \leq  - ( 280 \: \mbox{MeV} )^3 \label{range}
\label{qq6}
\eeq 
in order to include the central values and the errors
of the  QCD-SR and lattice estimates.

\subsection{The Constituent Quark Mass M}

Concerning the values of $M$,
 we take the point of view that it is  an arbitrary
parameter to be best fitted by comparing the predictions of the model with the 
experiments and by minimizing the theoretical $\gamma_5$-scheme dependence.
This point will be discussed further  at the end of this paper.

Let us only comment on what values we consider reasonable for $M$. 
As a general
rule, we expect $M$ to represent the constituent mass for the quarks inside
the Goldstone bosons (see \eq{CQM}),
and therefore to be roughly 
\beq
M  \approx 200-250 \: \mbox{MeV} \, ,
\eeq
as consistently estimated in 
processes~\cite{Bijnens} involving mesons. Such a value
 is smaller than the value  $M \simeq 330$ MeV
often
quoted that originates  
from baryon physics.

\section{The Matrix Elements}

We wish now to apply our results to kaon decays by computing the amplitudes
\beq
A_{00} \equiv A \left( K^0  \rightarrow \pi^0 \pi^0 \right) \, ,
\quad 
A_{+-} \equiv A \left( K^0  \rightarrow \pi^+ \pi^- \right)
\label{a00a+-}
\eeq
and
\beq
 A_{+0} \equiv A \left( K^+  \rightarrow \pi^+ \pi^0 \right)
\label{a+0}
\eeq
by means of which is possible to discuss, for instance, the $\Delta I = 1/2$
selection rule in kaon physics and $\varepsilon '/\varepsilon$.

\subsection{Chiral Loops and  Long-Distance Scale Dependence}

The long-distance scale dependence of the matrix elements is contained to
 this order in the one-loop
corrections induced by the propagation of
the mesons, a contribution that we have so far ignored. Their effect is
important and
essential in the matching procedure between Wilson coefficients and matrix
elements that
is central in our approach.

The computation is rather involved
because of the many terms;
the diagrams (see fig. 4)  include the vertex as well as the wave-function
one-loop renormalization (see ref.~\cite{PhD} for the Feynman rules and the
details
of the computation).

We give the final results for
the meson loop renormalization of the
matrix elements for the three processes in \eqs{a00a+-}{a+0}. 
The contribution to the tree-level
amplitude of each term of the chiral lagrangian is formally factorized out 
as a function of the input parameters, while the corresponding
loop renormalization is given in the form of numerical coefficients.
These numerical coefficients
are complicated functions of the masses and the coupling constant.
They are made of polynomial terms, generally
of the order of
 \beq
 \frac{m^2}{(4 \pi f)^2} \, ,
 \eeq
and logarithmic terms of the order of
 \beq
 \frac{m^2}{(4 \pi f)^2} \ln \frac{m^2_a}{m^2_b} \, ,
 \eeq
where the masses can be any among $m_\pi$, $m_K$ and $m_\eta$.
The values of the masses and other input variables are those given in
table 3 of appendix C.

Let us recall here that the leading-log approximation would
be particularly crude in
this case since
the large mass separation between $m_\eta$ and $m_\pi$ makes the
renormalization group scale particularly uncertain.

\begin{figure}[t]
\epsfxsize=14cm
\centerline{\epsfbox{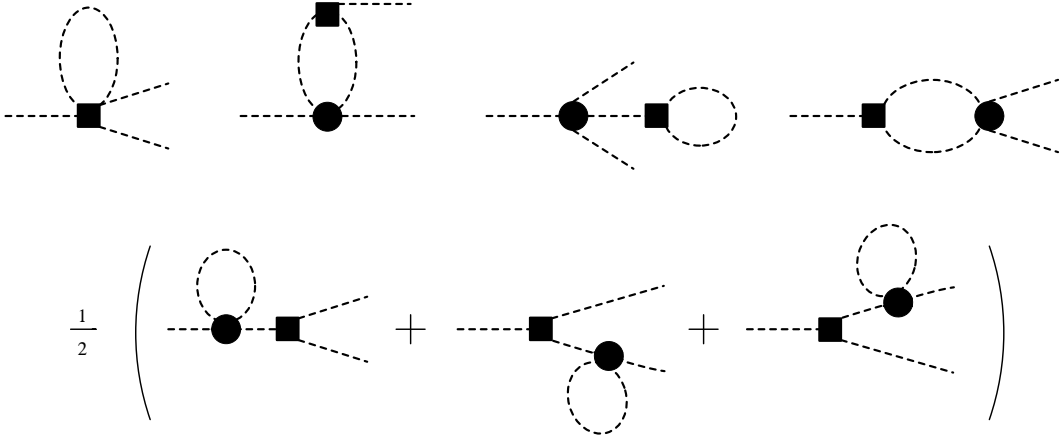}}
\caption{One-loop chiral renormalization of the the $K^0\to\pi\pi$ amplitudes.
The black box represents the insertion of the weak chiral lagrangian, whereas
the black circle denotes the insertion of the $O(p^2)$ strong chiral
lagrangian. For each chiral coefficient 
more than a hundred diagrams are generated by propagating
in all allowed ways $K$, $\pi$ and $\eta$. }
\end{figure}
The resulting corrected amplitudes for $K^0\to\pi^0\pi^0$ and
$K^0\to\pi^+\pi^-$ are given by:
\bea
a_{00} (Q_i) & = &  - \: \frac{\sqrt{2}}{f^3} \:
G_0 (Q_i) \left( 0.904 + 0.444 \, i + 0.255 \ln \ \mu^2 \right) \nnu \\
& & - \: \frac{\sqrt{2}}{f^3} (m_K^2 - m_\pi^2 ) \left[
G^a_{LL}(Q_i) \left(
0.367 + 0.444 \,i + 0.0747 \ln \ \mu^2 \right) \right. \nnu \\
& &\hspace*{2.5cm} \left . - \:G^b_{LL}(Q_i) \left(
1 + 0.349 + 0.0184 \,i + 0.135 \ln \ \mu^2 \right) \right. \nnu \\
& &\hspace*{2.5cm} \left. - \: G_{\underline{8}} (Q_i) \left(
1 + 0.716 + 0.463 \,i + 0.210 \ln \ \mu^2 \right) \right. \nnu \\
& &\hspace*{2.5cm} \left . + \: G^b_{LR}(Q_i) \left( 1 + 0.349 + 0.0184 \, i
 + 0.135 \ln \ \mu^2 \right)
\right]  \nnu \\
& & \left. - \frac{\sqrt{2}}{f^3} m_\pi^2 
\: G^a_{LR}(Q_i) \left(0.868 + 0.444 \, i  -0.132 \ln \ \mu^2 \right)
\right. \nnu \\
& & \left . + \frac{\sqrt{2}}{f^3} m_K^2 
\: G^c_{LR}(Q_i) \left(0.719 + 0.444 \, i + 0.203 \ln \ \mu^2 \right) \right.
 \label{00}
\eea
and
\bea
a_{+-} (Q_i) & = &  - \: \frac{\sqrt{2}}{f^3} \:
 G_0 (Q_i) \left( 1 + 0.708 + 0.240 \, i + 0.203 \ln
\ \mu^2 \right) \nnu \\
& &
- \frac{\sqrt{2}}{f^3} (m_K^2 - m_\pi^2 ) \left[
G^a_{LL}(Q_i) \left(
1 + 0.768 + 0.240 \,i + 0.278 \ln \ \mu^2 \right) \right. \nnu \\
& &\hspace*{2.5cm} \left . + \: G^b_{LL}(Q_i) \left(
0.0525 - 0.222 \,i + 0.0688 \ln \ \mu^2 \right)
\right. \nnu \\
& &\hspace*{2.5cm} \left. - \: G_{\underline{8}} (Q_i) \left(
1 + 0.716 + 0.463 \,i + 0.210 \ln \ \mu^2 \right) \right. \nnu \\
& &\hspace*{2.5cm} \left . - \: G^b_{LR}(Q_i) \left( 0.000127 - 0.222 \, i
 + 0.0583 \ln \ \mu^2 \right) 
\right]  \nnu \\
& & \left. - \frac{\sqrt{2}}{f^3} m_\pi^2  
\: G^a_{LR}(Q_i) \left( 1 + 4.17 + 0.240 \, i + 1.29 \ln \ \mu^2 \right)
\right. \nnu \\
& & \left. + \frac{\sqrt{2}}{f^3} m_K^2  
\: G^c_{LR}(Q_i) \left( 1 + 3.47 + 0.240 \, i + 1.40 \ln \ \mu^2 \right) 
\right. \, . 
\label{+-}
\eea
where all dimensionful parameters must be taken in units of GeV.

As a useful check we have also computed directly the meson loop
renormalization for the amplitude $K^+\to\pi^+\pi^0$, which is
a pure $\Delta I = 3/2$ transition:
\bea
a_{+0} (Q_i) & = & - \: \frac{1}{f^3} \:
G_0 (Q_i) \left( 1 - 0.196 - 0.204 \, i - 0.0510 \ln \ \mu^2 \right) \nnu \\
& &
-\frac{1}{f^3} (m_K^2 - m_\pi^2 ) \left[
G^{a}_{LL}(Q_i) \left(
 1   + 0.402
  - 0.204 \,i + 0.204 \ln \ \mu^2 \right) \right. \nnu \\
& &\hspace*{2.5cm} + \: G^{b}_{LL}(Q_i) \left(
 1   + 0.402
  - 0.204 \,i + 0.204 \ln \ \mu^2 \right)  \nnu \\
& &\hspace*{2.5cm} \left . - \: G^b_{LR}(Q_i) \left(1 + 0.349 - 0.204 \, i
 + 0.193 \ln \ \mu^2 \right) 
\right]  \nnu \\
& & \left.  - \frac{1}{f^3} m_\pi^2 
 \: G^a_{LR}(Q_i) \left( 1 + 3.30 - 0.204 \, i + 1.42 \ln \ \mu^2 \right)
\right. \nnu \\
& & \left.  + \frac{1}{f^3} m_K^2 
 \: G^c_{LR}(Q_i) \left(1 + 2.75 - 0.204 \, i +  1.20 \ln \ \mu^2 \right) 
\right.
\, .
\label{+0}
\eea

The renormalization at any scale $\mu$ can readily be obtained from
eqs. (\ref{00}) and (\ref{+-}).

As the reader can easily check, our result satisfies all the expected
symmetry properties, that is
\beq
A_{+-} = A_{00}
\label{test1}
\eeq
for the octet amplitudes, and
\beq
A_{+0} = \frac{A_{+-} - A_{00}}{\sqrt{2}}
\label{test2}
\eeq
for the other parts.

A similar computation for $G_{LL}^{a,b}$ and $G_{\underline{8}}$ is
reported in ref.~\cite{meson}.
Unfortunately, a comparison with  ref.~\cite{meson} is
difficult because the
authors do not report all the details of their computation. We find that
our imaginary parts are almost identical to theirs, while the real parts of
the renormalization
computed at the scale $m_\eta$, where the results of \cite{meson} are given,
differ, even though they are of the same order.
At any rate, the fact that \eqs{test1}{test2} are exactly satisfied for all
coefficients makes us confident on our results.

 Of course, the polynomial parts of these corrections receive
 contributions at the
 tree level from the next order terms in the chiral lagrangian. These are
 controlled by the two
 parameters $K_1$ and $K_4$ defined in  reference~\cite{meson}, where
 they are estimated to be
 \bea
 K_1 & = & (0.4 \pm 1.2) \times 10^{-11} \nnu \\
 K_4 & = & (0.3 \pm 1.4) \times 10^{-12} \, .
 \eea
 A similar result was obtained for $K_1$ in the framework of the $\chi$QM
 in~\cite{BEF}.

 These values correspond to a renormalization of the amplitudes
 $A_0$ and $A_2$ at the
percent level, which is an order of magnitude
smaller than the renormalization induced by chiral loops.
 Such a small contributions
are in agreement with the results of  the factorization and
 the vector dominance models, where these
 coefficients are vanishing. As a consequence, our results are not
 appreciably modified by the inclusion of these terms.

  We recall here that, contrary to the
 usual treatment, in our approach these next-to-leading-order
 contributions do not cancel the scale
 dependence of the chiral loops.

 On top of this renormalization, we should also include the one-loop
determination of
 $f$ in terms of $f_\pi$ and $f_K$. This is taken into account in 
 our computation by
replacing $f$ by
the one-loop renormalized value
 \beq
 f_1 = 0.087 \quad \mbox{GeV} \: \mbox{\cite{GL}}
 \eeq
in the tree level amplitudes. This correction amounts to a further 20 \% of
renormalization for the amplitudes.

Some of the loop renormalizations (the last two
terms in \eqs{00}{+0}) appear to be
large when compared to the partial tree level amplitudes that we have 
factorized out. This is a notational artifact since
they remain always
smaller than the leading tree level amplitude ($\sim \sqrt{2}/f^3$).
The overall renormalization is large but still under control; as
 a matter of fact, it is large in the $I=0$ channel and small in the
$I=2$ one (except for the subleading $LR$ momentum
corrections), thus  distributing itself in
the right amount to bring the matrix elements of the next subsection closer to
their experimental values.

In refs. \cite{cutoff,Paschos} a similar
computation was performed by means of a cut-off regularization, identifying
then the cut-off with the dimensional regularization scale
of the Wilson coefficients.
The two approaches lead to different results. Among else,
the meson loop renormalization of
the amplitude $A_2$ is strikingly different, being suppressed
in \cite{cutoff,Paschos}, whereas it is slightly enhanced in our approach.

\subsection{The Matrix Elements}

The  matrix elements of all the operators (\ref{Q1-10}) can now be computed.
By using the chiral lagrangian (\ref{chi-lag}), we can readily generate the
contributions with three mesonic external states necessary in the matrix
elements.
To these results we apply the one-loop meson renormalizations discussed in the
previous section, obtained from \eqs{00}{+-} by
subtracting the tree-level parts.

We write directly the expression for the isospin states
\beq
\langle Q_i \rangle_{0,2} \equiv \langle 2 \pi, I =0,2 | Q_i | K^0 \rangle \, ,
\eeq
Accordingly, the corresponding 
one-loop meson corrections are denoted by $a_{0,2}(Q_i)$.
The Clebsh-Gordan coefficients for the isospin projections 
can be found in the appendix. 

We recall that in the HV scheme one may rely 
on Fierz transformations and use the
factorized building blocks when computing the chiral coefficients. 
Fake HV anomalies must then be
subtracted according to the discussion in section 5.
Using the HV results of section 6 we obtain:
\bea
\langle Q_1 \rangle _0 & = & \frac{1}{3} X \left[ -1 + \frac{2}{N_c} \left(
1 - \delta_{\vev{GG}} \right)
\right] + a_0 (Q_1)\\
\langle Q_1 \rangle _2 & = & \frac{\sqrt{2}}{3} X \left[ 1 + \frac{1}{N_c}
\left(
1 - \delta_{\vev{GG}} \right)
\right] + a_2 (Q_1)\\
\langle Q_2 \rangle _0 & = & \frac{1}{3} X \left[ 2  - \frac{1}{N_c} \left(
1 - \delta_{\vev{GG}} \right)
\right] + a_0 (Q_2)\\
\langle Q_2 \rangle _2 & = &  \frac{\sqrt{2}}{3} X \left[ 1 + \frac{1}{N_c}
\left( 1 - \delta_{\vev{GG}} \right) \right] + a_2 (Q_2)\\
\langle Q_3 \rangle _0 & = & \frac{1}{N_c} X  \left(
1 - \delta_{\vev{GG}} \right) + a_0 (Q_3)\\
\langle Q_4 \rangle _0 & = & X + a_0 (Q_4) \\
\langle Q_5 \rangle _0 & = &  \frac{2}{N_c}  \, 
\frac{\langle \bar{q}q \rangle}{M f_\pi^2} \, X'  + a_0 (Q_5)\\
\langle Q_6 \rangle _0 & = &  2 \, 
\frac{\langle \bar{q}q \rangle}{M f_\pi^2} \, X' + a_0 (Q_6) \\
\langle Q_{7} \rangle _{0} & = & \frac{2 \sqrt{3}}{N_c} \, 
\frac{\langle \bar{q} q \rangle ^2}{f_\pi^3} 
- \frac{1}{N_c} \, 
\frac{\langle \bar{q}q \rangle}{M f_\pi^2} \, X'
  \nnu \\
 & & - \frac{2}{N_c} \frac{\vev{\bar{q}q}}{M f_\pi^2} \ Y
 + \frac{1}{2} X  + a_{0}(Q_7)\\
 \langle Q_{7} \rangle _{2} & = & \sqrt{6} \,
 \frac{\langle \bar{q} q \rangle ^2}{f_\pi^3} \frac{1}{N_c}
  - \frac{\sqrt{2}}{N_c} \frac{\vev{\bar{q}q}}{M f_\pi^2} \ Y
  - \frac{\sqrt{2}}{2} X + a_{2}(Q_7)\\
\langle Q_8 \rangle _0 & = &  2 \, \sqrt{3} \, 
\frac{\langle \bar{q} q \rangle ^2}{f_\pi^3} - 
\frac{\langle \bar{q}q \rangle}{M f_\pi^2} \, X' 
 \nnu \\
& & - 2 \, \frac{\vev{\bar{q}q}}{M f_\pi^2} \ Y
    + \frac{1}{2 N_c} X  \left(1 + \delta_{\vev{GG}}\right) + a_0(Q_8)\\
\langle Q_8 \rangle _2 & = &
\sqrt{6} \, \frac{\langle \bar{q} q \rangle ^2}{f_\pi^3}
  - \sqrt{2} \frac{\vev{\bar{q}q}}{M f_\pi^2} \ Y
  - \frac{\sqrt{2}}{2 N_c} X \left(1 + \delta_{\vev{GG}}\right)   + a_2(Q_8) \\
\langle Q_9 \rangle _0 & = &  - \frac{1}{2} X \left[ 1 - \frac{1}{N_c}
\left( 1 - \delta_{\vev{GG}} \right) \right] + a_0 (Q_9) \\
\langle Q_9 \rangle _2 & = &   \frac{\sqrt{2}}{2} X \left[ 1 + \frac{1}{N_c}
\left( 1 - \delta_{\vev{GG}} \right) \right] + a_2 (Q_9)\\
\langle Q_{10} \rangle _0 & = &   \frac{1}{2} X \left[ 1 - \frac{1}{N_c}
\left( 1 - \delta_{\vev{GG}} \right) \right] + a_0(Q_{10}) \\
\langle Q_{10} \rangle _2 & = &   \frac{\sqrt{2}}{2} X \left[ 1 + \frac{1}{N_c}
\left( 1 - \delta_{\vev{GG}} \right) \right] + a_2 (Q_{10})\, .
 \eea
where
\beq
X \equiv \sqrt{3} f_\pi \left( m_K^2 - m_\pi^2 \right)\ , 
\quad\quad 
X'  =   X \left( 1 - 6\,
\frac{M^2}{\Lambda_\chi^2} \right)
\eeq
and
\beq
Y \equiv \sqrt{3} f_\pi \left[m_\pi^2 + 3\ m_K^2
\frac{M^2}{\Lambda^2_\chi}\right] \ ;
\eeq
$\delta_{\vev{GG}}$ is given by (\ref{GG}).
 
  The one-loop 
renormalization of $f$  is taken into account by replacing
$f$ with $f_1$ in
the tree-level amplitudes, which amounts to replacing $1/f^3$ with $1/f^3_\pi$
multiplied by
\beq
1 + 3\ \frac{f_\pi - f_1}{f_\pi} \simeq 1.18\, .
\eeq

In the NDR scheme we are not allowed to Fierz transform the quark operators,
and we must resort to the direct computation of the unfactorized
two-loop diagrams. On the
other hand, we need not worry about chiral anomalies as in the HV scheme
and no subtraction is required. Using the NDR results
of section 6 we thus find:
 \bea
\langle Q_1 \rangle _0 & = & \frac{1}{3} X \left[ -1 + \frac{2}{N_c} \left(
1 - \delta_{\vev{GG}} \right)
\right] + a_0 (Q_1)\\
\langle Q_1 \rangle _2 & = & \frac{\sqrt{2}}{3} X \left[ 1 + \frac{1}{N_c}
\left(
1 - \delta_{\vev{GG}} \right)
\right] + a_2 (Q_1)\\
\langle Q_2 \rangle _0 & = & \frac{1}{3} X \left[ 2  - \frac{1}{N_c} \left(
1 - \delta_{\vev{GG}} \right)
\right] + a_0 (Q_2)\\
\langle Q_2 \rangle _2 & = &  \frac{\sqrt{2}}{3} X \left[ 1 + \frac{1}{N_c}
\left( 1 - \delta_{\vev{GG}} \right) \right] + a_2 (Q_2)\\
\langle Q_3 \rangle _0 & = & \frac{1}{N_c} \left( X'
-  \delta_{\vev{GG}} X \right) + a_0 (Q_3)
\\
\langle Q_4 \rangle _0 & = & X'  +a_0 (Q_4)\\
\langle Q_5 \rangle _0 & = & \frac{2}{N_c}  \,
\frac{\langle \bar{q}q \rangle}{M f_\pi^2} \, X''  + a_0 (Q_5)  \\
\langle Q_6 \rangle _0 & = & 2  \,
\frac{\langle \bar{q}q \rangle}{M f_\pi^2} \, X''  + a_0 (Q_6)  \\
\langle Q_{7} \rangle _{0} & = &  \frac{2 \sqrt{3}}{N_c} \, 
\frac{\langle \bar{q} q \rangle ^2}{f_\pi^3}
 \left( 1 - 3 \frac{M^3 f_\pi^2}{\langle \bar{q} q \rangle \Lambda_{\chi}^2}
 \right) - \frac{1}{N_c} \frac{\langle \bar{q}q \rangle}{M f_\pi^2} \, X'' 
  \nnu \\
 & & - \frac{2}{N_c} \frac{\vev{\bar{q}q}}{M f_\pi^2} \ Y'
 + \frac{1}{2} X + a_{0} (Q_7)\\
\langle Q_{7} \rangle _{2} & = &
\frac{1}{N_c} \sqrt{6} \, \frac{\langle \bar{q} q \rangle ^2}{f_\pi^3}
\left(1 - 3 \frac{M^3 f_\pi^2}{\langle \bar{q} q \rangle \Lambda_{\chi}^2}
\right)  \nnu \\
& &  - \frac{\sqrt{2}}{N_c} \frac{\vev{\bar{q}q}}{M f_\pi^2} \ Y'
- \frac{\sqrt{2}}{2} X + a_{2} (Q_7)\\
\langle Q_8 \rangle _0 & = &   2 \sqrt{3}\,
 \frac{\langle \bar{q} q \rangle ^2}{f_\pi^3}
 \left( 1 - 3 \frac{M^3 f_\pi^2}{\langle \bar{q} q \rangle \Lambda_{\chi}^2}
 \right) - \frac{\langle \bar{q}q \rangle}{M f_\pi^2} \, X'' 
 \nnu \\
& &  - 2 \, \frac{\vev{\bar{q}q}}{M f_\pi^2} \ Y' \:
+ \frac{1}{2 N_c} X \left(1 + \delta_{\vev{GG}}\right) + \: a_0(Q_8) \\
\langle Q_8 \rangle _2 & = &
\sqrt{6} \, \frac{\langle \bar{q} q \rangle ^2}{f_\pi^3}
\left(1 - 3 \frac{M^3 f_\pi^2}{\langle \bar{q} q \rangle
\Lambda_{\chi}^2} \right) \nnu \\
& &  - \sqrt{2} \frac{\vev{\bar{q}q}}{M f_\pi^2} \ Y'
 - \frac{\sqrt{2}}{2 N_c} X \left(1 + \delta_{\vev{GG}}\right)  \: +
a_2(Q_8)\\
\langle Q_9 \rangle _0 & = &  - \frac{1}{2} \left[ X - \frac{1}{N_c}
\left(2 X - X' - \delta_{\vev{GG}} X\right) \right] + a_0 (Q_9) \\
\langle Q_9 \rangle _2 & = &   \frac{\sqrt{2}}{2} X \left[ 1 + \frac{1}{N_c}
\left( 1 - \delta_{\vev{GG}} \right) \right] + a_2 (Q_9)\\
\langle Q_{10} \rangle _0 & = &   \frac{1}{2} \left[2 X - X' - \frac{1}{N_c}
\left(1 - \delta_{\vev{GG}}\right) X \right] + a_0(Q_{10}) \\
\langle Q_{10} \rangle _2 & = &   \frac{\sqrt{2}}{2} X \left[ 1 + \frac{1}{N_c}
\left( 1 - \delta_{\vev{GG}} \right) \right] + a_2 (Q_{10})\, .
\eea
where
\beq
X'' = X \left( 1 - 9\ \frac{M^2}{\Lambda_\chi^2} \right) \, ,
\quad \quad Y' \equiv \sqrt{3} f_\pi \left[m_\pi^2 + 3\ 
\left(m_K^2-m_\pi^2\right)
\frac{M^2}{\Lambda^2_\chi}\right]
\ .
\eeq

$\langle Q_i \rangle _2 = 0$ for $i = 3,4,5,6$ in both schemes.

The
equations above show the importance of the corrections of $O (\alpha_s  N)$
(parameterized by the value of the gluonic condensate) as well as of the
meson-loop
renormalizations. In the limit $\delta_{\vev{GG}}  \rightarrow 0$ and zero
meson-loop renormalization,
the HV hadronic matrix elements are the same as those
found in the $1/N_c$ approach, except for the $(V-A) \otimes (V+A)$
 operators  $Q_5$,
$Q_6$, $Q_7$ and $Q_8$  for which the detailed form of the matrix elements is
characteristic of the model employed. For instance, by means of \eq{qLqR}
we find 
\beq
\langle Q_6 \rangle _0  =  - 4  \,
\frac{\langle \bar{q}q \rangle ^2}{f_\pi^4 \Lambda_\chi^2} \, X \label{q6B}
\eeq
in the $1/N_c$ computation, where we used
\beq
c_1 + c_2 = \left(\frac{f_K}{f_\pi} -1 \right) \ \frac{\Lambda_\chi^2}
{m_K^2-m_\pi^2} \label{fk/fpi}
\eeq
 as determined from 
${\cal A}(K^+\to\mu^+\nu_\mu)/{\cal A}(\pi^+\to\mu^+\nu_\mu)$~\cite{1/N}. 
Eq.~(\ref{q6B}) shows the quadratic dependence
on the quark condensate which is distinctive of such an approach.   

To our knowledge the terms proportional
to $Y$ and $Y'$ in the matrix elements of $Q_7$ and $Q_8$ have 
been neglected so far. As an example, in the $1/N_c$
framework we find for the  matrix element 
$\langle Q_8 \rangle _2$:
\beq
\langle Q_8 \rangle _2 = 
\sqrt{6} \, \frac{\langle \bar{q} q \rangle ^2}{f_\pi^3}
  + 2 \sqrt{6} \, \frac{\vev{\bar{q}q} ^2}{f_\pi^3 \Lambda_\chi^2} 
  \left[ (c_1-c_2)\, m_\pi^2 - 
  {c_2} m_K^2 \right]
  - \frac{\sqrt{2}}{2 N_c} X 
\label{Q8Y} \, ,
\eeq
where the absolute values of $c_1$ and $c_2$ remain undetermined 
(only their sum is determined by \eq{fk/fpi}). 
Analogous contributions appear in the matrix elements of $Q_7$.
The matrix element (\ref{Q8Y})
correponds to the $1/N_c$ determination of the chiral coefficients
\beq
G^a_{LR} = - 6 \ \frac{\vev{\bar{q}q} ^2}{\Lambda_\chi^2}\: (c_1-c_2) \quad
\mbox{and} \quad G^c_{LR} = - 6 \ \frac{\vev{\bar{q}q} ^2}{\Lambda_\chi^2} 
\: c_2 \, ,
\eeq 
to be contrasted to that of the $\chi$QM in eqs. (\ref{gcHV}) and 
(\ref{gcNDR}).
In particular, the  $\chi$QM determination of $G^a_{LR}$ and $G^c_{LR}$
gives
$c_2/(c_1-c_2) \sim O(M^2/\Lambda_\chi^2)$, 
thus making the term proportional to
$m_K^2$ of the same order of magnitude of that proportional to $m_\pi^2$.

Although these additional contributions 
do not affect the estimate of the $\Delta I =1/2$ rule,
which is little sensitive to the electroweak penguins,
they do have an impact on the determination of 
$\varepsilon'/\varepsilon$~\cite{III}.

\subsection{A Final Comment}


In studying any physical process, we must consider
the matching of the hadronic matrix elements
with the Wilson coefficients  at a scale 
typically of the order of
$\Lambda_\chi$.
At the NLO, the Wilson coefficients of (\ref{ham}) depend---beside
the energy scale---on the
$\gamma_5$ scheme employed~\cite{Monaco,Roma}.
One should  thus verify to what extent the physical amplitudes
turn out to be renormalization scale independent, as a result of a balance
between short- and long-distance scale dependence.
On the other hand, one should  check the $\gamma_5$-scheme independence
of the results. The latter may allow us to restrict the 
range for the constituent quark mass $M$, as it was already
shown in ref. \cite{BEF} in a toy model for $\varepsilon'/\varepsilon$.

These issues deserve a separate and detailed analysis which is presented
in refs.~\cite{II} and~\cite{III}.
In the first of these two papers we address
the computation of the $\Delta I = 1/2$ rule in the $CP$-conserving kaon
decays. 
The results obtained are extremely encouraging, 
allowing us to give in~\cite{III} a new estimate  of the
$CP$-violating parameter $\varepsilon '/\varepsilon$
in the standard model.
A similar approach for the $\Delta S = 2$ lagrangian is also
under investigation \cite{A-L}.

\vspace*{1.5cm}

{\sc Acknowledgments}

\bigskip
We thank J. Gasser and R. Iengo for discussions.
M.F. thanks the Physics Department
at the University of Oslo for the hospitality.
This work was partially supported by the EEC Human Capital and
Mobility  contract ERBCHRX CT 930132.

\appendix
\section{Chiral Quark Model}
\subsection{Feynman Rules}

The free propagator for the constituent quark is given by
\beq
S_0 (p) = \frac{i}{\not p - M}\, ,
\eeq
where $\not p = \gamma \cdot p$.
The same propagator in the external gluon field (fixed-point gauge)
 is~\cite{S2}:
\beq
S_1 (p) = -\frac{i}{4}\ g_s  T^a G_{\mu\nu}^a\
\frac{R^{\mu\nu}}{(p^2 - M^2)^2} \, ,
\eeq
where
\beq
R^{\mu\nu} = \sigma^{\mu\nu}(\not p + M) +
(\not p + M)\sigma^{\mu\nu}
\eeq
and $\sigma_{\mu\nu} = (i/2) [\gamma_\mu ,\,
\gamma_\nu]$.

Other useful formul\ae\ are:
\beq
\Tr g_s^2 T^a T^b G_{\mu\nu}^a G_{\alpha\beta}^b =
\frac{\pi^2}{6} \langle \frac{\alpha_s}{\pi} GG \rangle \left(
\delta_{\mu\alpha}\delta_{\nu\beta} - \delta_{\mu\beta}\delta_{\nu\alpha}
\right) \, ,
\eeq
where $\Tr T^a T^b \equiv \Tr \lambda^a \lambda^b /4 = \delta^{ab}/2$, and
\beq
\sigma^{\mu\nu} \sigma_{\mu\nu} = 12\ \mbox{\bf I}\, ;
\qquad \sigma^{\mu\nu}
{\gamma}_\rho \sigma_{\mu\nu} = 0 \, .
\eeq

The relevant meson--quark interactions are derived from
the lagrangian (\ref{M-lag}), which  we write here as
\bea
{\cal L}_{\chi \mbox{\scriptsize QM}} & = & - M \bar{q} q + 2\, i\,
\frac{M}{f} \bar{q} \gamma_5 \, \Pi \, q
+ 2\, \frac{M}{f^2} \bar{q}\, \Pi^2 q \nnu \\
& & + \, \frac{4}{3} \, i \, \frac{M}{f^3} \bar{q} \gamma_5 \Pi^3 q + O (1/f^4
)
\, , \eea
where
\beq
\Pi = \frac{1}{2} \sum_a \lambda^a \pi^a = \frac{1}{\sqrt{2}}
\left[ \begin{array}{ccc} \tilde{\pi}^0 & \pi^+ & K^+ \\
                          \pi^- & -\bar{\pi}^0 & K^0 \\
                           K^- & \bar{K}^0 & \tilde{\pi}^8 \end{array}
\right]  \, ,
\eeq
and
\beq
\tilde{\pi}^0 = \frac{1}{\sqrt{2}} \pi^0 + \frac{1}{\sqrt{6}} \eta_8\, , \qquad
\bar{\pi}^0 = \frac{1}{\sqrt{2}} \pi^0 -
\frac{1}{\sqrt{6}} \eta_8\, , \qquad \tilde{\pi}^8 = -
\frac{2}{\sqrt{6}} \eta_8 \, .
\eeq
In the case of a single meson interactions one obtains
\beq
\bar{q} \gamma_5 \, \Pi \, q = \frac{1}{\sqrt{2}} \left( \bar{u} \gamma_5 u
\, \tilde{\pi}^0 - \bar{d} \gamma_5 d \, \bar{\pi}^0 + \bar{d} \gamma_5 s \,
K^0
+ \bar{u} \gamma_5 d \, \pi^+ + \cdots \: \right) \, .
\eeq
The relevant Feynman rules are therefore given by:
\bea
K^0\ \bar d\gamma_5 s  =  K^+\   \bar u\gamma_5 s  =
\pi^+\ \bar u\gamma_5 d -\mbox{coupling:}
&\qquad & - \frac{M\sqrt{2}}{f} \gamma_5 \nnu \\
 \pi^0\ \bar d\gamma_5 d -\mbox{coupling:}
&\qquad & + \frac{M}{f} \gamma_5 \nnu \\
\pi^0\ \bar u\gamma_5 u -\mbox{coupling:}
&\qquad & - \frac{M}{f} \gamma_5 \nnu \\
 K^0\ \pi^0\ \bar d s     -\mbox{coupling:}
&\qquad & - i\, \frac{M}{f^2\sqrt{2}}  \\
K^+\ \pi^-\ \bar d s = K^+\ K^-\ \bar u u   =
K^0\ \pi^+\ \bar u s -\mbox{coupling:}
&\qquad & + i\, \frac{M}{f^2} \nnu \\
 K^+\ K^-\ \bar s s  =  \pi^+\ \pi^-\ \bar u u  =
 \pi^+\ \pi^-\ \bar d d -\mbox{coupling:}
&\qquad & + i\, \frac{M}{f^2} \nnu \\
 K^+\ \pi^0\ \bar u s = \pi^+\ \pi^0\ \bar u d -\mbox{coupling:}
&\qquad & + i\,\frac{M}{f^2\sqrt{2}} \nnu \\
 \pi^0\ \pi^0\ \bar u u =   \pi^0\ \pi^0\ \bar d d
  -\mbox{coupling:}
&\qquad & + i\,\frac{M}{2 f^2} \nnu
\eea
All meson fields are entering the vertex.
The same rules hold for the conjugated couplings.

\subsection{Fierz Transformations and Clebsh-Gordan Coefficients}

The relevant Fierz transformations (for anti-commuting fields)
 are the following:
\bea
\bar{a}_\alpha \gamma_\mu ( 1 \pm \gamma_5 ) b_\beta \:
\bar{c}_\beta \gamma^\mu ( 1 \mp \gamma_5 ) d_\alpha
& = & - 2\, \bar{a}_\alpha  ( 1 \mp \gamma_5 ) d_\alpha \:
\bar{c}_\beta  ( 1 \pm \gamma_5 ) b_\beta \nnu \\
\bar{a}_\alpha \gamma_\mu ( 1 \pm \gamma_5 ) b_\beta \:
\bar{c}_\beta \gamma^\mu ( 1 \pm \gamma_5 ) d_\alpha
& = & \bar{a}_\alpha \gamma_\mu ( 1 \pm \gamma_5 ) d_\alpha \:
\bar{c}_\beta \gamma^\mu ( 1 \pm \gamma_5 ) b_\beta \, .
\label{diracF}
\eea
We also have for $SU(N_c)$
\beq
\delta_{\alpha \beta} \delta_{\gamma \delta} =
\frac{1}{N_c} \delta_{\alpha \delta} \delta_{\gamma \beta} +
2\ T^a_{\alpha \delta} T^a_{\gamma \beta} \, .
\label{colorF}
\eeq

The  $SU(2)$ Clebsh-Gordan projections are as given by
\bea
A_0 & = & \sqrt{\frac{1}{6}} \Bigl( A_{00} + 2 A_{+-} \Bigr) \nnu \\
A_2 & = & \sqrt{\frac{1}{3}} \Bigl( A_{+-} - A_{00} \Bigr) =
\sqrt{\frac{2}{3}} A_{+0}\, . \label{02}
\eea

The $SU(3)$ projections are~\cite{CG}
\bea
| \underline{8}, \,  \textstyle{\frac{1}{2}} \rangle & = & \: \Tr \left( \lambda^1_2
\Sigma^{\dag}
D_\mu  \Sigma \right)
\Tr \left(  \lambda^3_1 \Sigma^{\dag} D^\mu \Sigma \right) \nnu \\
& & - \:
\Tr \left( \lambda^3_2 \Sigma^{\dag} D_\mu
\Sigma \right)
\Tr \left(  \lambda^1_1 \Sigma^{\dag} D^\mu \Sigma \right)\nnu \\
| \underline{27}, \, \textstyle{\frac{1}{2}} \rangle & = &  \, \Tr \left( \lambda^1_2
\Sigma^{\dag}
D_\mu  \Sigma \right)
\Tr \left(  \lambda^3_1 \Sigma^{\dag} D^\mu \Sigma \right) \nnu \\
& & + \: 4 \, \Tr \left( \lambda^3_2 \Sigma^{\dag} D_\mu
\Sigma \right)
\Tr \left(  \lambda^1_1 \Sigma^{\dag} D^\mu \Sigma \right) \nnu \\
& & + \: 5  \, \Tr \left( \lambda^3_2 \Sigma^{\dag} D_\mu
\Sigma \right)
\Tr \left(  \lambda^2_2 \Sigma^{\dag} D^\mu \Sigma \right)\nnu \\
| \underline{27}, \, \textstyle{\frac{3}{2}} \rangle & = &  \, \Tr \left( \lambda^1_2
\Sigma^{\dag}
D_\mu  \Sigma \right)
\Tr \left(  \lambda^3_1 \Sigma^{\dag} D^\mu \Sigma \right) \nnu \\
& & + \: \Tr \left( \lambda^3_2 \Sigma^{\dag} D_\mu
\Sigma \right)
\Tr \left(  \lambda^1_1 \Sigma^{\dag} D^\mu \Sigma \right) \nnu \\
& & -  \: \Tr \left( \lambda^3_2 \Sigma^{\dag} D_\mu
\Sigma \right)
\Tr \left(  \lambda^2_2 \Sigma^{\dag} D^\mu \Sigma \right) \, .
\eea
Therefore, we have
\bea
| \underline{27} \rangle & = & \frac{5}{9} \: | 
\underline{27}, \, {\textstyle{\frac{3}{2}}} \rangle +
\frac{1}{9} \: | \underline{27}, \, \textstyle{\frac{1}{2}} \rangle \nnu \\
& = & \frac{2}{3}\,  \Tr \left( \lambda^1_2 \Sigma^{\dag}
D_\mu  \Sigma \right)
\Tr \left(  \lambda^3_1 \Sigma^{\dag} D^\mu \Sigma \right) \nnu \\
 & & + \:  \Tr \left( \lambda^3_2 \Sigma^{\dag} D_\mu
\Sigma \right)
\Tr \left(  \lambda^1_1 \Sigma^{\dag} D^\mu \Sigma \right) \, .
\eea

\subsection{Dimensional regularization}

We work in the $\overline{\mbox{MS}}$ scheme.
In the naive dimensional regularization (NDR) everything is continued to $d$
dimensions and the same four-dimensional rules applied. We therefore have that
\beq
\{ \gamma_\mu ,\, \gamma_\nu \} = 2\, g_{\mu \nu}
\eeq
and
\beq
g^\mu_\mu = d \, .
\eeq

The $\gamma_5$ matrix is defined so as to anti-commute in any dimensions as
\beq
 \{ \gamma_\mu ,\, \gamma_5 \} = 0 \, .
 \eeq

 The term naive refers to the fact that such a prescription leads to manifest
algebraic inconsistencies.

In the 't Hooft-Veltman regularization (HV) the Dirac matrices are separately
considered in 4 (tilded quantities) and $d-4$ (hatted quantities) dimensions
so that
\beq
\gamma_\mu = \tilde{\gamma}_\mu + \hat{\gamma}_\mu \, .
\eeq
The two sub-spaces are orthogonal to each other:
\beq
\{ \tilde{\gamma}_\mu ,\, \tilde{\gamma}_\nu \} = 2\, \tilde{g}_{\mu \nu} \quad
 \{ \hat{\gamma}_\mu ,\, \hat{\gamma}_\nu \} = 2\, \hat{g}_{\mu \nu} \quad
\{ \hat{\gamma}_\mu ,\, \tilde{\gamma}_\nu \} = 0 \, ,
\eeq
and
\beq
\hat{g}^\mu_\mu = d - 4 \quad \tilde{g}^\mu_\mu = 4 \quad \mbox{and}
\quad
\hat{g}^\mu_\alpha \tilde{g}^\alpha_\nu = 0 \, .
\eeq

The $\gamma_5$ matrix is defined as anti-commuting  in 4 dimensions
and commuting in $d-4$; therefore
\beq
\{ \tilde{\gamma}_\mu ,\, \gamma_5 \} =0 \quad
[ \hat{\gamma}_\mu ,\, \gamma_5 ] = 0 \, .
\eeq

The rules above lead to
\beq
\{\gamma^\mu ,\: \gamma_5  \} =
 2 \, \gamma_5 \hat{\gamma}^\mu
\, .
\eeq

In both schemes the external momenta are kept in four dimensions. Chiral
currents must be symmetrized in order to have a unique definition. This is
immaterial in the NDR case but gives
\beq
\frac{1}{2} ( 1 + \gamma_5 ) \gamma_\mu ( 1 - \gamma_5 ) = \tilde{\gamma}_\mu
( 1 - \gamma_5 )
\eeq
in the HV case.

\subsection{Trace Formul\ae}

In writing the chiral $\Delta S =1$
lagrangian we have rewritten single traces as 
the product of two traces. The trace factorization properties
can be easily shown. 
The eight $SU(3)$ Gell-Mann matrices 
$\lambda^a$ together with the identity matrix 
$\lambda^0$ form a basis
for the $3 \times 3$ complex matrices with non-vanishing trace.
The standard normalization
$\Tr \lambda^a \lambda^b = 2 \delta^{ab}$ implies
\beq
\lambda^0 \equiv \sqrt{\frac{2}{3}} \hat{I} \, ,
\label{norm}
\eeq
Given two complex $3 \times 3$ matrices $A$ and $B$
we can write
\bea
\lambda^i_j A &=& \sum_{a=0}^8 \lambda^a a^a \nnu \\
\lambda^m_n B &=& \sum_{a=0}^8 \lambda^b b^b \, , 
\label{expan}
\eea
where $a^a$ and $b^b$ are complex numbers. From \eqs{norm}{expan} 
it follows that
\beq
\Tr \left(  \lambda^i_j A\ \lambda^m_n B \right) = \frac{1}{2}\ \sum_{a=0}^8
\Tr \left( \lambda^a \lambda^i_j A \right)\
\Tr \left( \lambda^a \lambda^m_n B \right) \, .
\eeq
Finally, by using the explicit form of the Gell-Mann matrices and the identity,
we have, for the cases of interest, 
\bea
\Tr \left(  \lambda^3_2 A\ \lambda^1_1 B \right) & = &
\Tr \left(\lambda^1_2 A \right)\
\Tr \left( \lambda^3_1 B \right) \\
\Tr \left(  \lambda^3_1 A\ \lambda^1_2 B \right) & = &
\Tr \left(\lambda^1_1 A \right)\
\Tr \left( \lambda^3_2 B \right) \, .
\eea

\newpage

\section{Chiral Perturbation Theory: Feynman Rules}

We give in the two following tables the chiral perturbation theory rules needed
in computing the coefficients of the weak chiral lagrangian.
Many more rules are necessary in the chiral loop computation (see~\cite{PhD}).

\begin{table}[h]
\begin{center}
\begin{tabular}{|c||c|c|}
\hline
 & \multicolumn{2}{c|}{ \rm Feynman Rule } \\
\hline
 {\rm Coefficient} & $K^0 (p_1)\pi^0 (p_2)$ & $K^+ (p_1)\pi^- (p_2)$ \\
 \hline
$G^{(0)}$  & 0 & $  i \frac{2}{f^2}  $  \\
$G_{\underline{8}}~~ $    & $i \frac{\sqrt{2}}{f^2}\: p_1 \cdot p_2$  &
                                     $-i \frac{2}{f^2}\: p_1 \cdot p_2 $\\
$G^a_{LL}$ & 0  & $i \frac{2}{f^2}\: p_1 \cdot p_2$ \\
$G^b_{LL}$ & $i \frac{\sqrt{2}}{f^2}\: p_1 \cdot p_2 $  & 0 \\
$G^a_{LR}$ &  0 &  $- i \frac{2}{f^2}\: p_1 \cdot p_2$ \\
$G^b_{LR}$ & $- i \frac{\sqrt{2}}{f^2}\: p_1 \cdot p_2$   & 0 \\
$G^c_{LR}$ & 0 & 0 \\
\hline
\end{tabular}
\end{center}
\end{table}
\begin{table}[h]
\begin{center}
\begin{tabular}{|c||c|c|}
\hline
 & \multicolumn{2}{c|}{ \rm Feynman Rule } \\
\hline
 {\rm Coefficient} & $K^0 (p_1)\pi^0 (p_2) \pi^0 (p_3)$ &
 $K^0 (p_1)\pi^+ (p_2) \pi^- (p_3)$ \\
 \hline
$G^{(0)}$  & 0 & $  - \frac{\sqrt{2}}{f^3} $  \\
$G_{\underline{8}}~~ $    & $ - \frac{1}{\sqrt{2}f^3} \left[ \frac{1}{2} p_1
\cdot ( p_2  + p_3)
                       - p_2 \cdot p_3 \right] $  &
                         $\frac{\sqrt{2}}{f^3} \: p_2 \cdot ( p_3 - p_1 ) $\\
$G^a_{LL}$ & 0  & $ -\frac{\sqrt{2}}{f^3} \:p_3 \cdot ( p_2 - p_1 ) $ \\
$G^b_{LL}$ & $ - \frac{1}{\sqrt{2}f^3} \left[ \frac{1}{2} p_1 \cdot ( p_2  +
p_3)
                       - p_2 \cdot p_3 \right] $  &
                        $-  \frac{\sqrt{2}}{f^3}\: p_1 \cdot ( p_2 - p_3 )$ \\
$G^a_{LR}$ &  0 & $  \frac{\sqrt{2}}{f^3} \:p_3 \cdot ( p_1 + p_2 )$ \\
$G^b_{LR}$ & $ \frac{1}{\sqrt{2}f^3} \left[ \frac{1}{2} p_1 \cdot ( p_2  + p_3)
                       - p_2 \cdot p_3 \right]$   &
                       $-  \frac{\sqrt{2}}{f^3} \: p_1 \cdot ( p_2 - p_3 )$ \\
$G^c_{LR}$ & 0 & $ -\frac{2 \sqrt{2}}{f^3} \:p_1 \cdot  p_2 $ \\
\hline
\end{tabular}
\end{center}
\caption{Relevant Feynman rules of chiral perturbation theory.
All momenta are entering the vertex.}
\end{table}

\newpage

\section{Input Parameters}

\begin{table}[ht]
\begin{center}
\begin{tabular}{|c|c|}
\hline
{\rm parameter} & {\rm value} \\
\hline
$f_\pi = f_{\pi^+}$  &  92.4  MeV \\
$f_K = f_{K^+}$ & 113 MeV \\
$m_\pi = (m_{\pi^+} + m_{\pi^0})/2 $ & 138 MeV \\
$m_K = m_{K^0}$ &  498 MeV \\
$m_\eta$ & 548 MeV \\
\hline
$\Lambda_{QCD}^{(4)}$ & $350 \pm 100$ MeV \\
$\overline{m}_u + \overline{m}_d$ (1 GeV) & $12 \pm 2.5$ MeV \\
$\vev{\bar{q}q}$  &  $- \mbox{(200 -- 280 MeV)} ^3$ \\
$ \langle \alpha_s G G/\pi \rangle $ & $(376 \pm  47 \:
\mbox{MeV} )^4 $ \\
\hline
\end{tabular}
\end{center}
\caption{Table of the numerical values used for the input parameters.}
\end{table}
%

\clearpage
\renewcommand{\baselinestretch}{1}

\end{document}